\documentclass[twocolumn]{aastex63}
\usepackage{amsmath}

\newcommand{\rp}{r_{\rm p}}

\renewcommand{\vec}[1]{\boldsymbol{#1}}

\newcommand{\rrel}{r_{\rm rel}}
\newcommand{\rcap}{r_{\rm cap}}
\newcommand{\RH}{R_{\rm H}}
\newcommand{\RHm}{R_{\rm H,mut}}
\newcommand{\RHot}{R_{\rm H12}}

\newcommand{\be}{\begin{eqnarray}}
\newcommand{\ee}{\end{eqnarray}}

\newcommand\js{\bgroup\markoverwith{\textcolor[rgb]{1.0, .5, .1}{\rule[0.5ex]{8pt}{1.5pt}}}\ULon}

\submitjournal{ApJ} 

\shorttitle{BBH close encounters}
\shortauthors{Li, Lai, and Rodet}

\graphicspath{{./}{figures-1020/}}

\begin{document}

\title{Long-term Evolution of Tightly-Packed Stellar Black Holes in AGN Disks:
Formation of Merging Black-Hole Binaries via Close Encounters
}

\correspondingauthor{Jiaru Li}
\email{jiaru\textunderscore li@astro.cornell.edu}

\author[0000-0001-5550-7421]{Jiaru Li}
\affiliation{Center for Astrophysics and Planetary Science,
Department of Astronomy, Cornell University, Ithaca, NY 14853, USA}

\author[0000-0002-1934-6250]{Dong Lai}
\affiliation{Center for Astrophysics and Planetary Science,
Department of Astronomy, Cornell University, Ithaca, NY 14853, USA}

\author[0000-0002-1259-3312]{Laetitia Rodet}
\affiliation{Center for Astrophysics and Planetary Science,
Department of Astronomy, Cornell University, Ithaca, NY 14853, USA}

\begin{abstract}
We study the long-term evolution of two or more stellar black holes (BHs) on initially separated but unstable circular orbits around a supermassive BH (SMBH).
Such a close-packed orbital configuration can naturally arise from BH migrations in the AGN disk. 
Dynamical instability of the orbits leads to recurring close encounters between two BHs, during which the BH separation $\rp$ becomes less than the Hill radius $\RH$.  
In the rare very close encounters (with $\rp$ several orders of magnitude less than $\RH$), a tight merging BH binary can form with the help of gravitational wave emission.  
We use $N$-body simulations to study the time evolution of close encounters of various degrees of ``closeness'' and the property of the resulting binary BH mergers.  
For a typical ``SMBH + 2 BHs'' system, the averaged cumulative number of close encounters (with $\rp \lesssim \RH$) scales approximately as $\propto t^{0.5}$.  
The minimum encounter separation $\rp$ follows a linear cumulative distribution $P(<\rp) \propto \rp$ for $\rp \ll \RH$.  
From these, we obtain a semi-analytical expression for the averaged rate of binary captures that lead to BH mergers.
Our results suggest that close-packed BHs in AGN disks may take a long time ($\gtrsim 10^7$ orbits around the SMBH) to experience a sufficiently close encounter and form a bound binary, although this time can be shorter if the initial BH orbits are highly aligned.
The BH binary mergers produced in this scenario always have high eccentricities when entering the LIGO band, and have a broad distribution of orbital inclinations relative to the original AGN disk.
We also explore the effects of the gas disk and find that simple gas drags on the BHs do not necessarily lead to an enhanced BH binary capture rate.
\end{abstract}

\keywords{Active galactic nuclei(16); 
Black holes(162);
Galaxy accretion disks(562);
Gravitational wave sources(677);
N-body simulations(1083)
}

\section{Introduction}  
\label{sec:intro}

Binary black hole (BBH) mergers are the most common sources of gravitational waves (GWs) detected by the LIGO/Virgo collaboration \citep{LIGO2019, LIGO2021a, LIGO2021b}.
A variety of formation channels for these merging BBHs have been studied, ranging from isolated binary evolution \citep[e.g.,][]{Lipunov1997, Podsiadlowski2003, Belczynski2010, Belczynski2016, Mandel2016, Marchant2016}, strong gravitational scatterings in dense star clusters \citep[e.g.,][]{PortegiesZwart2000, OLeary2006, Miller2009, Banerjee2010, Downing2010, Ziosi2014, Samsing2014, Rodriguez2015, Samsing2018, Kremer2019}, and tertiary-induced mergers in stellar triple/quadrupole systems \citep[e.g.,][]{Miller2002, Silsbee2017, Liu2018, Liu2019, Liu2019a, Fragione2019} or in nuclear clusters around a central supermassive BHs
\citep[e.g.,][]{Antonini2012, VanLandingham2016, Petrovich2017, Hamers2018, Liu2019b, Liu2021}.
The probability and significance of each channel are still unclear.

The possibility of BBH mergers in AGN accretion disks around supermassive black holes (SMBHs) has received much attention in recent years \citep[e.g.,][]{McKernan2012, McKernan2014, Bartos2017, Stone2017, Secunda2019, Secunda2020, Yang2019a, Yang2019b, Grobner2020, Ishibashi2020, Tagawa2020b, Tagawa2020a, Ford2021}.
BBHs in flat disks may be hardened, or even driven to merger, by a series of nearly co-planar binary-single scatterings \citep[e.g.,][]{Stone2017, Leigh2018, Samsing2020}.
Hydrodynamical interaction between the gaseous AGN disk and a pre-existing BH binary may also influence the orbital evolution of the binary \citep{Baruteau2011, Li2021a, Li2021, Li2022}.
BH Mergers that happen inside AGN disks could have several observable properties, such as the possible associations with electromagnetic counterparts \citep{Stone2017, McKernan2019, Graham2020} and distinct mass and spin distributions \citep[e.g.,][]{McKernan2018, Yang2019a}.

The AGN disk channel of BBH mergers typically relies on the disks being an ideal environment for dynamical formation of BH binaries.
BBHs could be ``pre-existing'' in nuclear star clusters and get captured in the inner AGN disks \citep{Bartos2017} or form in situ in the extended region of the disks \citep{Stone2017}.
It was suggested that AGN disks \citep{Sirko2003, Thompson2005} may contain migration traps where stellar-mass BHs (sBHs) can accumulate \citep{Bellovary2016}. Orbiters inside such traps can have close encounters with each other due to their mutual gravity and potentially form binaries \citep{Secunda2019, Secunda2020}.  
A direct gas-capture channel has also been proposed to form binaries from single sBHs \citep{Tagawa2020a}.

An important and unsolved issue of the AGN disk BBH merger scenario is the evolution of sBHs during close encounters and how bound BBHs actually form.  
Previous studies tend to adopt very simple prescriptions for BBH formation and follow-up evolution.
For example, \cite{Secunda2019} performed $N$-body simulations of multiple sBHs and included disk force prescriptions to mimic the effects of eccentricity damping and migration traps, but they assumed that all close encounters between two BHs within the mutual Hill radius and with negative relative binary energy lead to the formation of bound binaries that quickly merge. 
This assumption is problematic as the vast majority of such bound binaries are short-lived and are quickly destroyed by the tidal force of the SMBH. 
\cite{Tagawa2020a} considered gas-assisted binary formation in their population synthesis study, including the time delay between BBH formation and merger, and allowing newly formed BBHs to be disrupted before mergers.
However, their studies are one-dimensional and they assumed that the relative orbits of all BBHs are circular for simplicity. 

In this paper, we study how often two or more sBHs in closely-packed, initially circular and nearly co-planar orbits around a SMBH can be captured into a ``permanent'' binary and merge with the aid of the gravitational radiation. 
Such close-packed orbits (with the difference in orbital radii less than a few times the Hill radius $\RH$) can be naturally produced by the differential migrations of sBHs in the AGN disk and/or the migration traps \citep{Bellovary2016, Secunda2019}.
Since the sBHs in close orbits are dynamically unstable, they exhibit chaotic orbital motion around the SBMH and undergo repeated close encounters with each other (with their separation less than $\RH$).  
Binary capture occurs only in the very rare occasion when two sBHs experience an extreme close encounter (with their separation several orders of magnitude less than $\RH$), during which energy dissipation through GW emission is effective.
We perform a large number of $N$-body simulations to study the occurrence rate of such close encounters and the properties of the captured BBHs.
Since the dynamical influence of the disk gas on sBHs during close encounters is highly uncertain, the major part of our paper focuses on the clean problem where the only dissipative effect is GW emission.
Nevertheless, we also carry out an exploratory study on the gas effects by adding
``frictional'' forces on the sBHs to mimic the BH-gas interactions. 

The rest of this paper is structured as follows.  
In Section~\ref{sec:GWCE}, we introduce our scenario for BBH formation in AGN disks via close encounters between sBHs.  
In Section~\ref{sec:fiducial}, we describe our fiducial ``SMBH + 2 BHs'' simulations with no gas effects included.  
We use these simulations to obtain the time evolution of the close encounter rate, the distribution of the minimum sBH separations during encounters, and the timescale (and the probability) to form long-lived or merging BH binaries. 
Section~\ref{sec:inc} discusses how our results depend on the initial inclinations of the BH orbits around the SMBH.
In Section~\ref{sec:friction} we apply simple models of the disk forces on the sBHs to assess the effects of gas disks on the evolution of the embedded sBHs and the formation rate of merging BBHs.
Section~\ref{sec:NBH} explores the rate and properties of the close encounters in systems with more than two sBHs around the SMBH.
In Section~\ref{sec:summary}, we summarize our findings.

\section{Scenario}
\label{sec:GWCE}

AGN disks can help bringing stellar-mass black holes (sBHs) circulating around a supermassive BH (SMBH) into close orbits due to the differential migrations of the BHs and migration traps \citep{Bellovary2016}, therefore promoting close encounters between the sBHs. 
While the encountering sBHs typically have too much relative kinetic energy to become bound in the presence of the tidal field of the SMBH, they may occasionally have a very close encounter, during which gravitational wave (GW) emission can take away the excessive energy. 
Dynamical friction from the disk gas may also play a role, but its effect is more difficult to quantify.
These very close encounters may turn the two BHs into a bound binary and lead to BBH merger.

We consider a system with a central SMBH of mass $M$, around which orbit two or more sBHs on nearly circular and nearly co-planar trajectories. 
For simplicity, henceforth `BHs' always refer to stellar-mass black holes that are orbiting around the SMBH.
Due to their migrations in the AGN disk, the BHs may have orbits very close to each other.
For the most part of this paper, we set up two BHs with masses $m_1$, $m_2$ and initial semi-major axes $a_1$, $a_2$ around the SMBH.
If the dimensionless orbital separation (in units of the mutual Hill radius $\RHot$) is less than a critial value, i.e.
\be
\label{eq:a1-a2-criterion}
\frac{a_2-a_1}{\RHot} \equiv K < K_{\rm crit} \sim 1,
\ee
where 
\be
\label{eq:RH}
\RHot \equiv \frac{a_1+a_2}{2} \left(\frac{m_1+m_2}{3M}\right)^{1/3},
\ee
the orbits are dynamically unstable, such that the two BHs will soon develop orbital crossing and start chaotic evolution. 
The boundary between ``stable'' and ``unstable'' can be fuzzy but the critical value $K_{\rm crit}$ is of order unity and depends on the ``frictional'' force acting on the BH from the disk gas (Li, Rodet and Lai 2022).
In the absence of BH-disk interaction, the Hill stability criterion gives $K_{\rm crit}=2\sqrt{3}$ \citep{Gladman1993}.

There are three possible outcomes for the two BHs in unstable orbits: 
(i) The lighter BH ($m_2$) is ejected from the system;
(ii) The two BHs experience a sufficiently close encounter such that GW emission and/or gas drag makes them into a bound binary and eventually merge; 
(iii) One of the BHs moves very close to the SMBH and gets ``swallowed'' by it.
Outcome (iii) has a negligible probability when $a_1,a_2 \gg GM/c^2$, i.e. when the horizon radius of the SMBH is much less than the BH orbital distances.
In our ``SMBH + 2 BHs'' systems with $M \gg m_1, m_2$, outcome (i) will take many orbital periods to happen. 
This can be understood as follows.
Close encounters between $m_1$ and $m_2$ ($<m_1$) cause $m_2$ to experience energy diffusion, with the change (loss or gain) of energy during each encounter given by $\Delta E \sim \alpha (G m_1 m_2 / a_1)$, where $\alpha\gtrsim1$.
Thus the average number of close encounters required for $m_2$ to be ejected is $\left< N_{\rm ej} \right> \sim (G M m_2/2a_1)^{2}(\Delta E)^{-2} \sim (4\alpha^2)^{-1}(M/m_1)^2$. Indeed, extensive numerical experiments carried out by \cite{Pu2021} show that $N_{\rm ej}$ has a broad distribution, with the mean value given by (see their Eq 24)
\be
\left< N_{\rm ej} \right> \simeq 0.06^2 \left( \frac{M}{m_1} \right)^2 \left(1+\frac{m_2}{m_1}\right)^4,
\ee
and the distribution has a long tail at the larger values (the $68\%$ quantile of $N_{\rm ej}$ ranges from $0.25\left< N_{\rm ej} \right>$ to $13\left< N_{\rm ej} \right>$). The ejection time is usually much longer then $N_{\rm ej}P_2$ (where $P_{2}$ is the initial orbital period of $m_{2}$) since the semi-major axis of $m_2$ increases as it approaches ejection. Thus, for $M/m_1 \gtrsim 10^{6}$, the ejection timescale $t_{\rm ej} \gg 10^{10}P_2$, which means ejection almost never happens.

We are thus left with outcome (ii), i.e. BH binary formation due to dissipative processes.
If we neglect the possible effect of gas drag, the only dissipation is GW emission (``gravitational bremsstrahlung''). 
During a very close encounter (i.e. the separation between $m_1$ and $m_2$ is much less than $\RH$), the two BHs lose their relative energy by the amount \citep{Peters1964,Quinlan1989}
\be
\label{eq:dEGW}
\Delta E_{\rm GW} = \frac{85\pi}{12\sqrt{2}}\frac{G^{7/2} \mu^2 m_{12}^{5/2}}{c^5 \rp^{7/2}},
\ee
where $m_{12} = m_1+m_2$, $\mu=m_1m_2/m_{12}$, and $\rp$ is the periastron separation of the relative trajectory of the BHs. 
To form a binary that is stable in the presence of the SMBH tidal field, we require 
\be
\Delta E_{\rm GW} \gtrsim \eta\frac{G m_1 m_2}{\RHot},
\ee
with $\eta$ of order unity. This implies that the pericenter distance of the relative orbit of $m_1$ and $m_2$ must be less than the critical value for capture, given by
\be
\nonumber
\frac{\rcap}{\RHot} 
& \simeq &
2.85 \eta^{-2/7} \left(\frac{\mu}{m_{12}}\right)^{2/7} 
\left(\frac{m_{12}}{M}\right)^{10/21}
\left(\frac{a_{12}}{GM/c^2}\right)^{-5/7}, \\
\label{eq:GWCE-rp_crit}
& \simeq &
10^{-4} \eta^{-2/7} \left(\frac{4\mu}{m_{12}}\right)^{2/7} 
\left(\frac{10^{6}m_{12}}{M}\right)^{10/21} \\
\nonumber && \times
\left(\frac{a_{12}}{100GM/c^2}\right)^{-5/7}, 
\ee
where $a_{12}=(a_1+a_2)/2\simeq a_1$. Thus, for the two unstable BHs to be captured into a bound binary requires an extremely close encounter with $\rp\lesssim10^{-4}\RHot$. 
A major goal of our paper is to evaluate the rate of such extremely close encounters and the properties of the resulted merging BH binaries.

\section{Formation of Bound Binaries in ``SMBH + 2 BHs'' Systems}
\label{sec:fiducial}

In this section, we study the formation rate of bound binaries for two BHs in unstable orbits around a SMBH. We neglect the effects of gas disk (if any) here -- these effects will be studied in Section~\ref{sec:friction}.  

\subsection{Setup of simulations}
\label{sec:fiducial-setup}

The fiducial system of our simulations consists of a central SMBH with mass $M$ and two smaller BHs with mass $m_1=2\times10^{-5}M$ and $m_2=10^{-5}M$. The initial orbital separation is set as
\be
a_2-a_1 = 2 \RHot,
\ee
so that their orbits are unstable. 
For convenience, henceforth we define 
\be
R_{\rm H} \equiv a_1 \left(\frac{m_1}{3M}\right)^{1/3} = \left(\frac{m_1}{m_{12}}\right)^{1/3} \RHot
\ee
to be the Hill radius of the $m_1$ at the beginning of the simulation and use it as the unit of distance.
We let $P_1$ be the initial orbital period of $m_1$ and use it as the unit of time. 
We note that $P_1$ at $a_1=100GM/c^2$ for $M=10^6 M_{\odot}$ is $10^{-3}$ yr. 
The BHs are given initial eccentricities $e_1=0$, $e_2=10^{-5}$, and inclinations $i_1=i_2=R_{\rm H}/a_1$. 
We carry out 2000 runs and sample the initial values of the argument of the peripasis, the longitude of the ascending node, and the mean anomaly randomly in the range $[0,2\pi]$ for each BH, assuming they all have uniform distributions.

We simulate the evolution of the system using the $N$-body code \textsc{REBOUND} \citep{Rein2012} with the \textsc{IAS15} integrator \citep{Rein2015}. Each simulation runs for $10^{5}$ $P_1$. 

\subsection{Close encounters}
\label{sec:fiducial-CE}

To characterize encounters with various degrees of ``closeness'', we designate a close encounter event to be ``CE$\alpha$'' when the separation between the BHs, $\rrel = \left|\vec{r}_1 - \vec{r}_2\right|$, becomes less than $10^{-\alpha}\RH$ and numerically keep track of CE0, CE1 and CE2 in each simulation. 
A CE$\alpha$ event starts when $\rrel$ changes from $>10^{-\alpha}\RH$ to $<10^{-\alpha}\RH$, and ends when $\rrel$ becomes greater than $10^{-\alpha}\RH$ again and the relative energy $E_{\rm rel}=\frac{1}{2}\mu\left|\vec{v}_1 - \vec{v}_2\right|^2 - \frac{Gm_1m_2}{\rrel}$ is positive.
The whole process is regarded as a single CE$\alpha$ event no matter how long it elapses.
In the simulation, a new CE$\alpha$ is logged only if the previous one has ended. 

\begin{figure}[t]
    \epsscale{1.2}
    \plotone{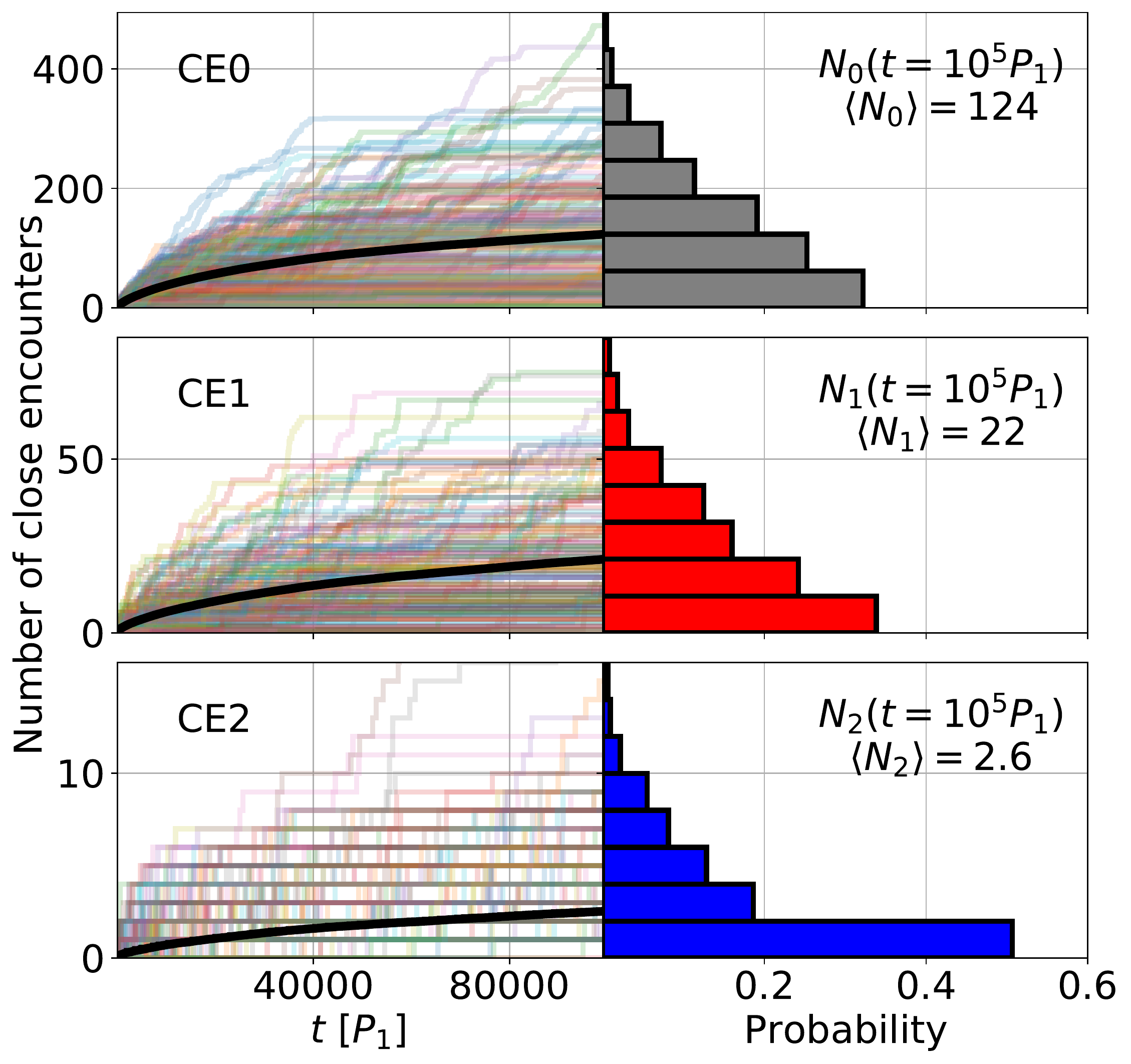}
    \caption{{\bf Left:} Cumulative number of CE0, CE1 and CE2 events as a function of time for our fiducial ``SMBH+2BHs'' systems. 
    A CE$\alpha$ event occurs when the separation between the two BHs becomes less than $10^{-\alpha}\RH$. 
    In each panel, each of the 200 colored curves represents a simulation of the ``SMBH+2BHs'' system. The black curve shows the average number of CE$\alpha$ events (averaged over all 2000 runs). 
    {\bf Right:} Distribution of $N_\alpha$ at $t=10^5P_1$ from all simulations. 
    The horizontal axis shows the probability of each bin.
    }
    \label{fig:NCE-vs-t}
\end{figure}

The left panels of Figure~\ref{fig:NCE-vs-t} show the time evolution of the CE0, CE1, and CE2 counts in 200 examples (out of a total of 2000) of our simulations. 
Because of the stochastic nature of the evolution, the cumulative event count, $N_{\alpha}(t)$, in one simulation can differ significantly from another. 
The right panel of Figure~\ref{fig:NCE-vs-t} shows the probability distribution of $N_{\alpha}(t=10^{5}P_1)$ from all 2000 simulations.
The total numbers of CE0, CE1, CE2 events in those simulations are 248790, 43153, 5238, respectively. 
Thus, at $t=10^{5}P_1$, the average number per simulation of CE0, CE1, and CE2 are $\left<N_{\alpha}(10^{5}P_1)\right> = $ 124, 22, and 2.6, respectively.
Note that $N_{\alpha}$ has a wide distribution: For example, while $\left<N_{2}(10^{5}P_1)\right> = 2.6$, $5\%$ of the runs have $N_{2}(10^{5}P_1)>10$.

Despite the difference in the numbers, the three kinds of CEs all have higher occurrence rate at the early times than at the later times. 
The time evolution of the average can be described by a power law $\langle N_{\alpha}(t) \rangle \propto t^{n_{\alpha}}$.
We perform least-square fits using such power-law to the results at $t\gtrsim2\times10^4P_1$ to exclude the initial transient stage.
We find
\be
\label{eq:N-vs-t}
\langle N_{0}(t) \rangle  & = & 0.67\left(\frac{t}{P_1}\right)^{0.45},\\
\langle N_{1}(t) \rangle  & = & 0.067\left(\frac{t}{P_1}\right)^{0.50},\\
\label{eq:N-vs-t-fid}
\langle N_{2}(t) \rangle  & = & 0.0064\left(\frac{t}{P_1}\right)^{0.52}.
\ee

\subsection{Very close encounters and BBH formation rate}
\label{sec:fiducial-rp}

During each CE$\alpha$ event, we take a simulation snapshot every $10^{-3}P_1$  and monitor the separation $\rrel = |\vec{r}_1 - \vec{r}_2|$ between the two BHs. 
The exact periapsis passage moment may lie between two of the snapshots. 
We use the snapshot right after the true pericenter passage (when $\rrel$ first increases) to calculate the relative energy and angular momentum of the encountering BHs:
\be
E_{\rm rel} & = & \frac{1}{2}\mu \vec{v}_{\rm rel}^2 - \frac{Gm_1m_2}{\rrel}, \\
\vec{\ell}_{\rm rel} & = & \vec{r}_{\rm rel} \times \vec{v}_{\rm rel}.
\ee
We then compute the semi-major axis $a_{\rm rel}$ of the orbit via $E_{\rm rel} = -Gm_1m_2/(2a_{\rm rel})$, and the pericenter distance $\rp$ using $\ell_{\rm rel}=\sqrt{Gm_{12}a_{\rm rel}(1-e_{\rm rel}^2)}\simeq\sqrt{2Gm_{12}\rp}$.

\begin{figure}[ht]
    \epsscale{1.0}
    \plotone{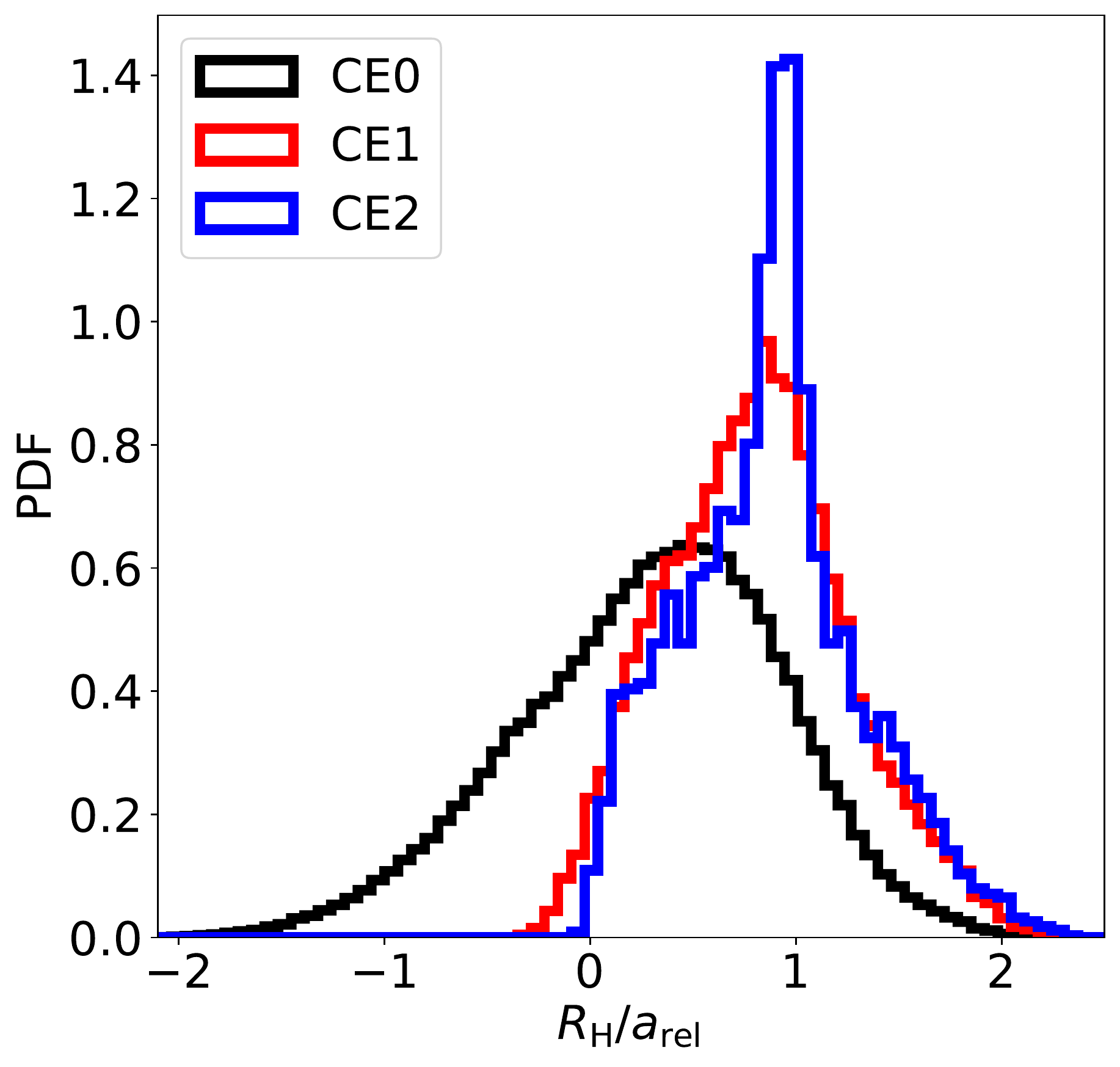}
    \caption{Probability distribution of $\RH/a_{\rm rel}$ in our fiducial simulations, where $a_{\rm rel}$ is the semi-major axis of the relative orbit of the BH pairs undergoing close encounters. 
    }
    \label{fig:PDF-aCE}
\end{figure}

Figure~\ref{fig:PDF-aCE} shows the distribution of the semi-major axis $a_{\rm rel}$ of the BH binary undergoing close encounters. 
The most likely $a_{\rm rel}$ is about $\RH$, and nearly all close encounters have $\RH/a_{\rm rel}\leq2.3$. 
Given that the BH pairs during the encounters are either unbound or have $a_{\rm rel} \gtrsim \RH$, they quickly depart from each other after reaching their minimum separation and get disrupted by the SMBH tide. 
This implies that the lifetime of most binaries formed by CEs is less than their orbital period, which is approximately equal to their period around the SMBH when $a_{\rm rel}\sim\RH$. Indeed, we find that, among the 248790 CE0 events in our 2000 simulations, the vast majority ($96.35\%$) disentangle within one orbital period ($P_1$) around the SMBH, and only $0.02\%$ survive for more than $10P_1$.

\begin{figure}[ht]
    \epsscale{1.0}
    \plotone{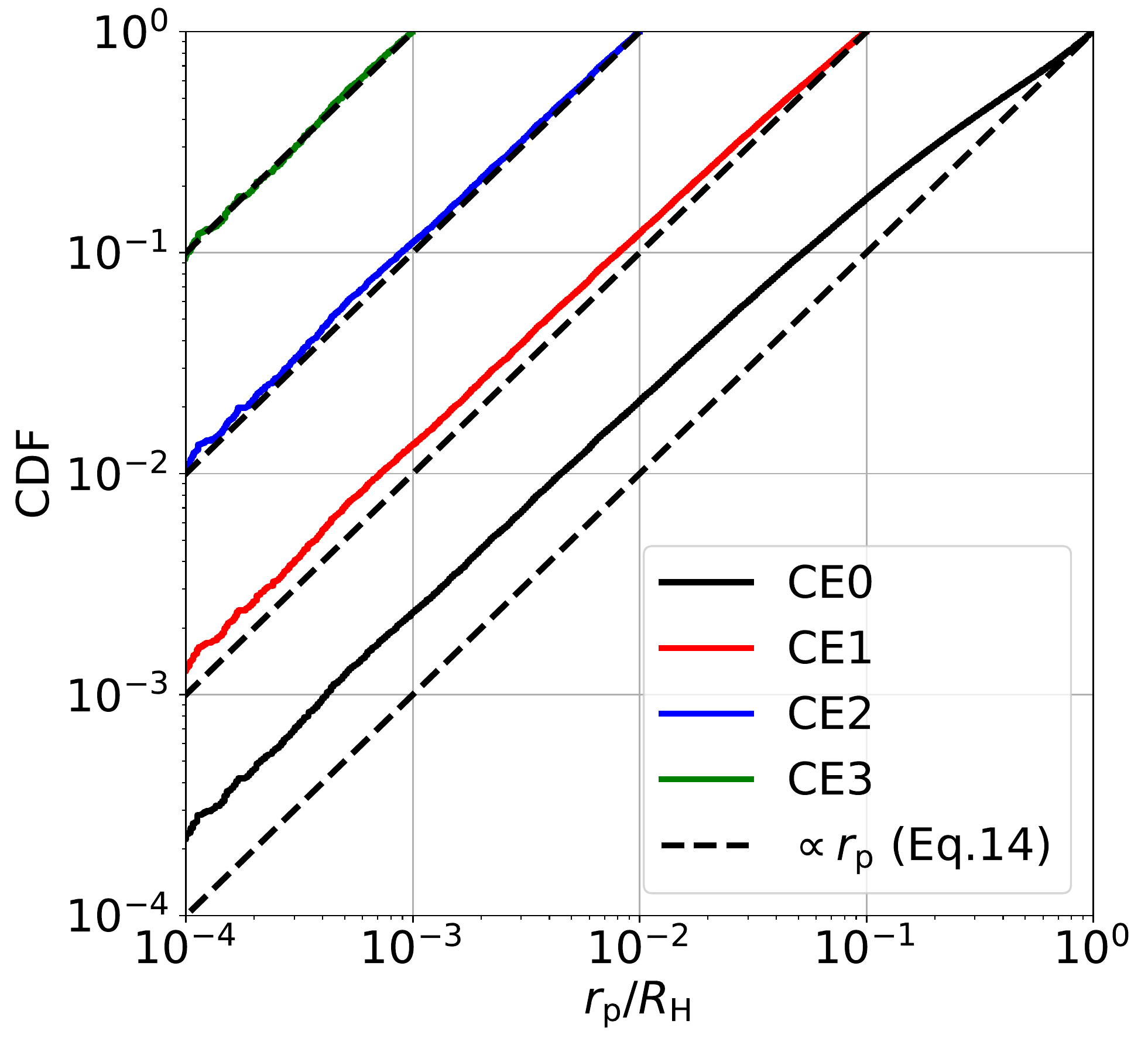}
    \caption{Cumulative distribution of $\rp$ in close encounters in our fiducial simulations. The solid colored curves show the distribution normalized for CE0, CE1, CE2 and CE3. The dashed black lines are given by equation~\eqref{eq:Probability-vs-rp}.
    }
    \label{fig:CDF-rp}
\end{figure}

Figure~\ref{fig:CDF-rp} shows the distribution of the pericenter (``closest'') distance of the two BHs during close encounters. We find that, for a CE$\alpha$ event, the cumulative distribution of $\rp$ (i.e., the probability for the pericenter distance to be less than $\rp$) is approximately given by 
\be
\label{eq:Probability-vs-rp}
P_{\alpha}(<\rp) \simeq \frac{\rp}{10^{-\alpha}\RH} \qquad (\text{for }\rp\leq10^{-\alpha}\RH).
\ee
Equation~\eqref{eq:Probability-vs-rp} becomes increasingly accurate for closer encounters. 
The difference in $\langle N_{\alpha}(t) \rangle$ for CE0, CE1 and CE2 discussed in Section~\ref{sec:fiducial-CE} is a direct consequence of this cumulative distribution of $\rp$. 

Equation~\eqref{eq:Probability-vs-rp} is equivalent to a constant probability distribution $dP_{\alpha}/d\rp = \text{const}. $ Since the relative angular momentum is $\ell_{\rm rel} \simeq \sqrt{2Gm_{12}\rp}$ for nearly parabolic encounters (for $\rp \ll a_{\rm rel} \sim \RH$; see Figure~\ref{fig:PDF-aCE}), this is equivalent to the probability distribution in $\ell_{\rm rel}$ given by 
\be
\label{eq:PDF-vs-l}
\frac{dP_{\alpha}}{d\ell_{\rm rel}} \propto \ell_{\rm rel}.
\ee
This relation has been previously obtained both analytically and numerically in the context of planetary collisions \citep{Li2020}. 
It can be understood as follows: When the two BHs reach a close separation $\rrel\ll\RH$, the angular momentum is $\ell_{\rm rel}=v_{\rm rel}r_{\perp}$, where $v_{\rm rel} \simeq \sqrt{2Gm_{12}/\rrel}$ and $r_{\perp}$ is the projection of $\vec{r}_{\rm rel}$ perpendicular to $\vec{v}_{\rm rel}$; 
if the initial mutual inclination of the two BHs is much larger than $\rrel/a_1$, the vector $\vec{r}_{\perp}$ is sampled uniformly in the plane perpendicular to $\vec{v}_{\rm rel}$ \citep{Li2020}, i.e. $dP/(r_{\perp}dr_{\perp})=\text{const}$, which implies Equation~\eqref{eq:PDF-vs-l}.

\begin{figure}[ht]
    \epsscale{1.0}
    \plotone{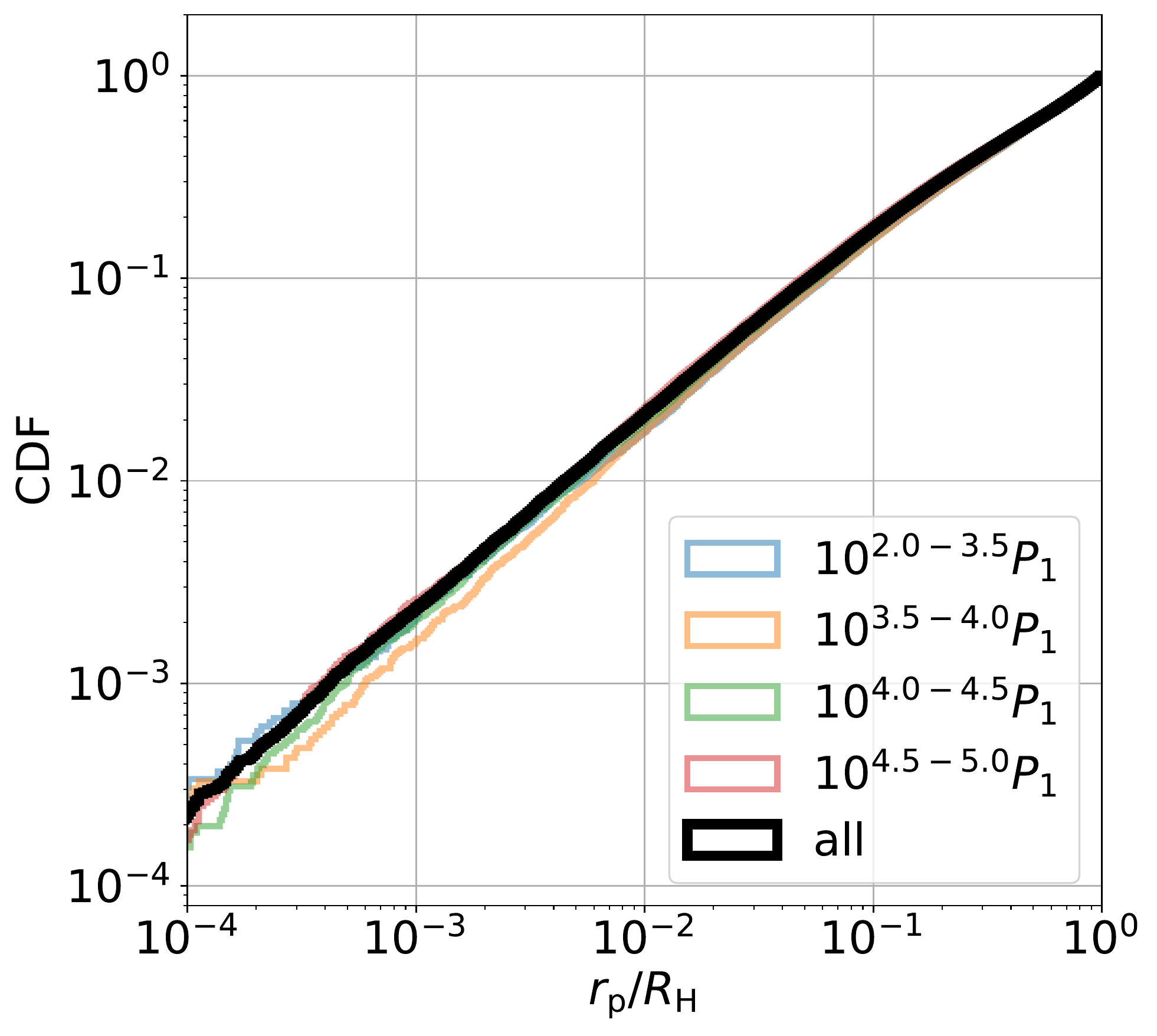}
    \caption{Cumulative distribution of $\rp$ in close encounters (CE0s) detected at different time intervals (as labeled) in our fiducial simulations. The black line here is the same as the black line in Figure~\ref{fig:CDF-rp}. 
    }
    \label{fig:CDF-rp-time}
\end{figure}

While Figure~\ref{fig:CDF-rp} depicts the $\rp$-distribution during the whole simulation time span, Figure~\ref{fig:CDF-rp-time} shows that the CEs from different time intervals also follow to the same distribution.

Thus, combining Equations~\eqref{eq:N-vs-t} and~\eqref{eq:Probability-vs-rp}, the averaged cumulative number of very close encounters with $\rp<\rcap$ given by
\be
\label{eq:N-rcap-t}
\langle N(t; \rp<\rcap) \rangle 
= \langle N_{\alpha}(t)\rangle P_{\alpha}(<\rcap).
\ee
Evaluating Equation~\eqref{eq:N-rcap-t} using $\left< N_2(t) \right>$ and $P_2(<\rcap)$, we find
\be
\label{eq:Nprod}
\langle N(t; \rp < \rcap) \rangle
&\simeq& 6\times10^{-5} \left(\frac{t}{P_1}\right)^{0.52} \nonumber \\
&& \times \left( \frac{\rcap}{10^{-4}\RH} \right),
\ee
where we have scaled $\rcap$ according to the estimate given by Equation~\eqref{eq:GWCE-rp_crit}. Thus, for $\rcap \simeq 10^{-4}\RH$, it would take about $10^{8}P_1$ on average for the two BHs to capture into a merging binary by GW emission.
Note that, because of the wide spread of $N(t)$ (see Figure~\ref{fig:NCE-vs-t}), it is possible for an unstable pair of BHs to reach $N(t)\simeq 4\left<N(t)\right>$ (about $5\%$ of the systems) and be captured in $\sim10^7P_1$.

\begin{figure}[ht]
    \epsscale{1.0}
    \plotone{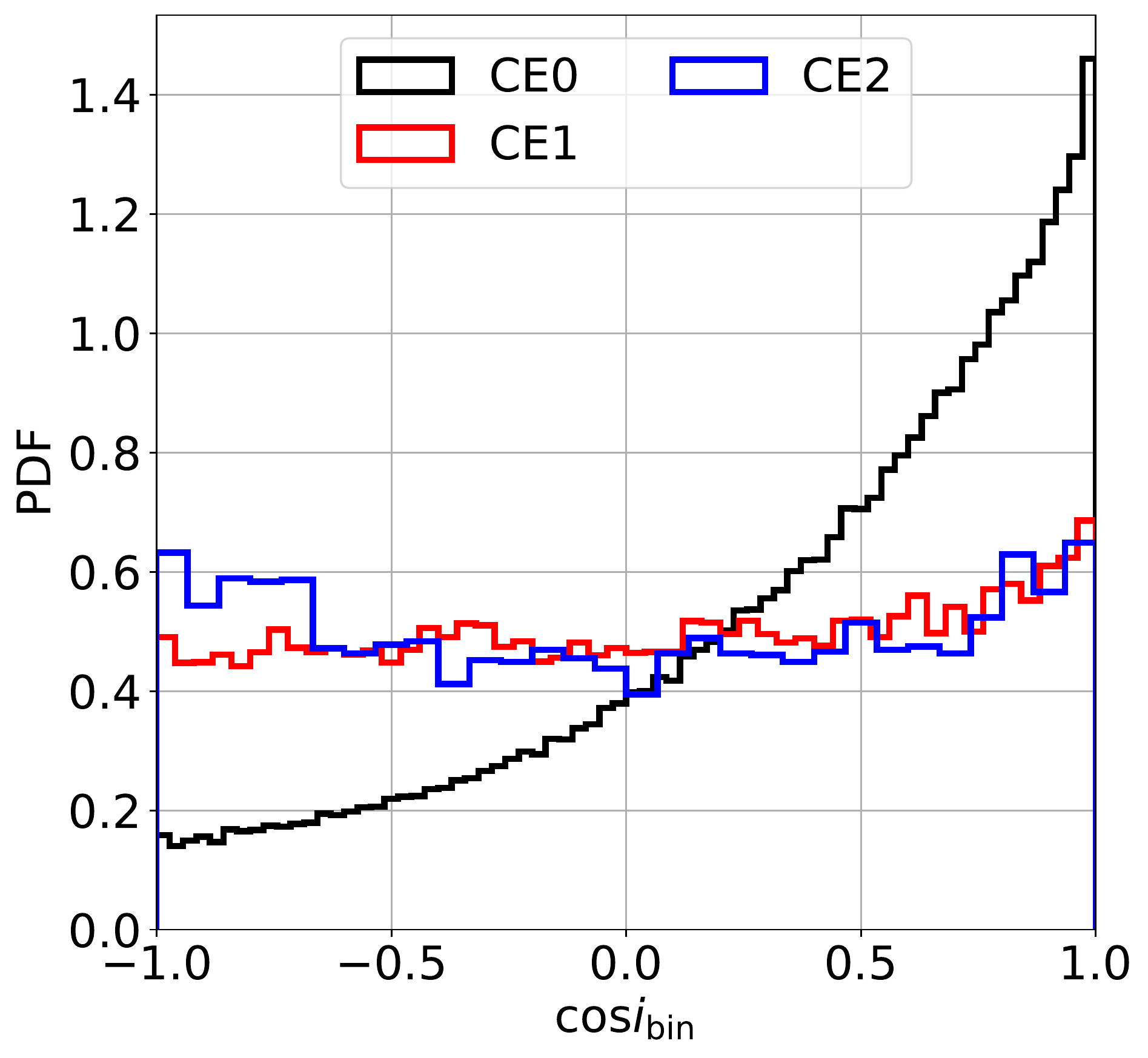}
    \caption{Probability density function of the binary obliquity $i_{\rm bin}$ of BHs in close encounters. Note that $i_{\rm bin}$ measures the angle between the close encounter plane and the orbital plane around the SMBH, i.e. $\cos{i_{\rm bin}}\equiv\hat{\ell}_{\rm rel}\cdot\hat{z}$, where $\hat{\ell}_{\rm rel}$ is the relative angular momentum axis of the two BHs, and $\hat{z}$ is the initial orbital axis around the SMBH. 
    }
    \label{fig:PDF-inc}
\end{figure}

\subsection{Orbital obliquities of BH binaries formed in close encounters}

The orbital obliquities ($i_{\rm bin}$) of the BH binaries formed by CEs can be inferred from the inclination of the angular momentum vector $\vec{\ell}_{\rm rel}$ of the CE orbits with respect to the initial orbital axis ($\hat{z}$) around the SMBH, i.e. $\cos{i_{\rm bin}}=\hat{\ell}_{\rm rel}\cdot\hat{z}$. 
Figure~\ref{fig:PDF-inc} shows the distribution of $\cos{i_{\rm bin}}$ for the close encounters in our fiducial simulations.
Encounters within Hill radius (CE0, black) are predominantly prograde with the most probable $i_{\rm bin}$ being zero. 
But closer encounters (CE1 and CE2) have nearly uniform distribution of $\cos{i_{\rm bin}}$. 
This indicates that merging BH binaries formed in very close encounters ($\rp\ll\RH$) have a broad range of inclinations with respect to the AGN disk plane.

The broad distribution of $\cos{i_{\rm bin}}$ is consistent with the findings of \cite{Li2020} in the context of planet collisions. 
Such a broad distribution arises when the initial mutual inclination of the BH orbits $\Delta i$ is much larger than $\rp/a_1$. Since our simulations have $\Delta i \sim \RH/a_1$, all CEs with $\rp\ll\RH$ will have a broad $\cos{i_{\rm bin}}$ distribution.
\footnote{\cite{Li2020} showed that when $\Delta i \gg \rp/a_1$ and $\Delta i \ll \RH/a_1$, the distribution of $\cos{i_{\rm bin}}$ is $f(\cos{i_{\rm bin}})\propto (1-\cos^2{i_{\rm bin}})^{-1/2}$, and that when $\Delta i \gg \rp/a_1$ and $\Delta i \sim \RH/a_1$, $f(\cos{i_{\rm bin}})$ becomes more uniform.}

\begin{figure}[t]
    \epsscale{1.2}
    \plotone{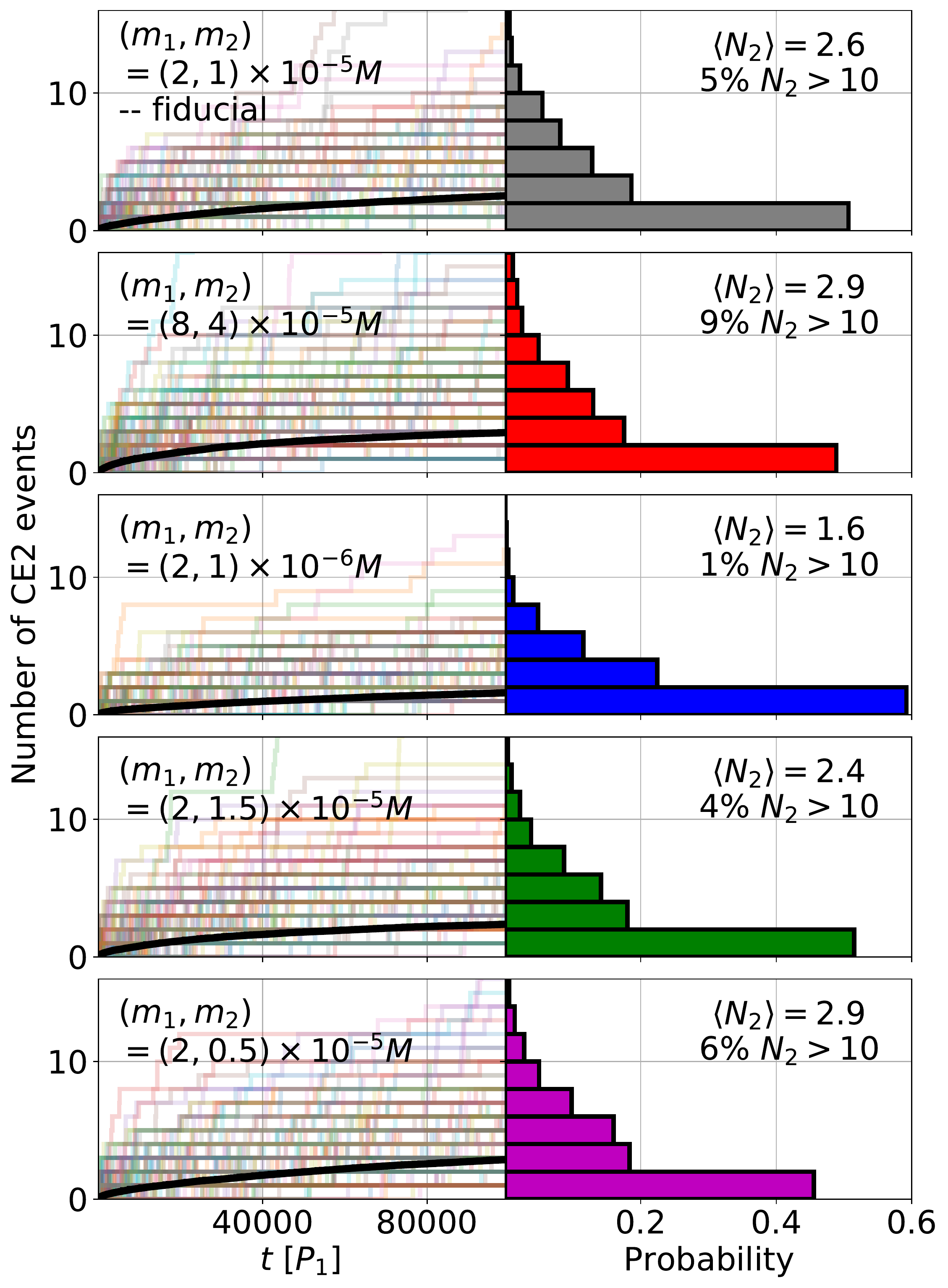}
    \caption{Same as Figure~\ref{fig:NCE-vs-t}, except showing the number of CE2 events in the systems with different BHs masses (as labeled).
    The texts on the right panels show the average number of $N_2$ at $t=10^{5}P_1$ and the percentage of the runs that experience more than 10 CE2 events.
    }
    \label{fig:NCE-vs-t-3v}
\end{figure}

\subsection{Mass dependence of the close encounter rate}
\label{sec:fid-mass}

The mass ratio between the SMBH and the BHs in AGN disks can span several orders of magnitude. 
We repeat our fiducial experiment (which has $m_1=2m_2=2\times10^{-5}M$) with four different sets of BH masses: $(m_1,m_2)=(8,4)\times10^{-5}M$,  $(m_1,m_2)=(2,1)\times10^{-6}M$, $(m_1,m_2)=(2,1.5)\times10^{-5}M$, and $(m_1,m_2)=(2,0.5)\times10^{-5}M$. 
We use them to test the effect of having larger BH mass, smaller BH mass, smaller BH mass ratio ($m_1/m_2=4/3$), and larger BH mass ratio ($m_1/m_2=4$), respectively.

Figure~\ref{fig:NCE-vs-t-3v} compares the CE2 event rates in the fiducial system and in the four systems with different BH masses.
In all cases, the $N_2$-distributions at $t=10^5P_1$ show large spreads, with the mean $\langle N_2(t=10^5P_1) \rangle$ in the range of $1.6$-$2.9$ (see Figure~\ref{fig:NCE-vs-t-3v}).
The probabilities for high $N_2$s (beyond the mean) are more different: for example, the probability for each simulation to get more than 10 CE2 events is $5\%$, $9\%$, $1\%$, $4\%$, and $6\%$ for the five cases displayed in Figure~\ref{fig:NCE-vs-t-3v}.

We find that, for four systems with different BH masses, the probability distributions of $\RH/a_{\rm rel}$, $\rp/\RH$ and $i_{\rm bin}$ are all similar to the fiducial results as depicted in Figures~\ref{fig:PDF-aCE},~\ref{fig:CDF-rp} and~\ref{fig:PDF-inc}. 
 
Overall, our results suggest that the BBH formation rate obtained for our fiducial system in Section~\ref{sec:fiducial-rp} can be applied to other systems with different BH masses (as long as $m_1, m_2 \ll M$) to within a factor of 2. 
The biggest effect that we observe is a drop in the CE2 rate by a factor of 2 when the BH masses are lowered by a factor of 10 compared to the fiducial system.

\begin{figure}[ht]
    \epsscale{1.0}
    \plotone{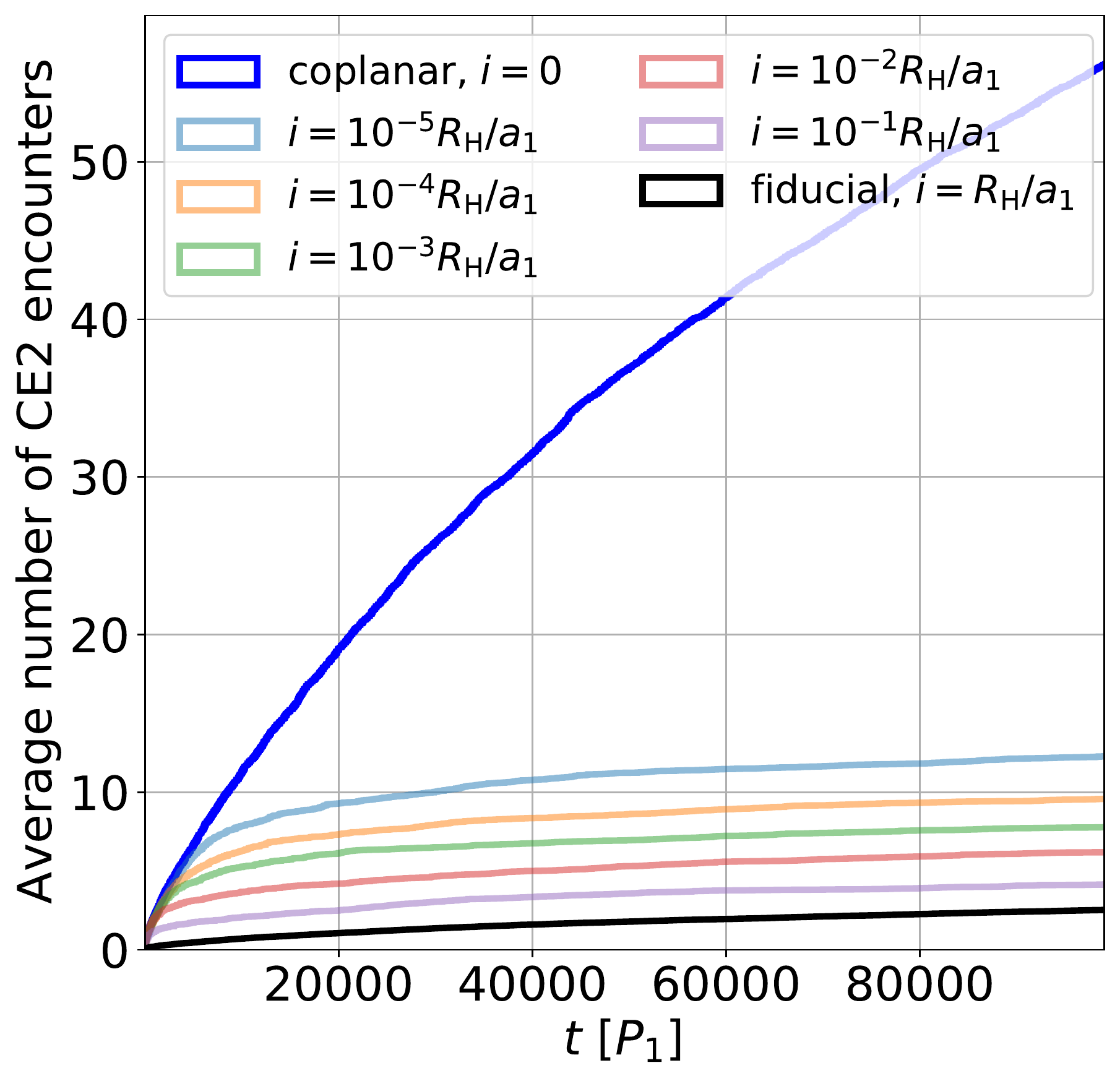}
    \caption{Average number of CE2 events as a function of time in systems with different initial orbital inclinations ($i_1=i_2=i$ as labeled). The black curve is the same as the black curve in the bottom left panel of Figure~\ref{fig:NCE-vs-t}.
    }
    \label{fig:NCE-vs-t-all-inc}
\end{figure}

\section{Systems of different initial inclinations}
\label{sec:inc}

\subsection{Results with different initial inclinations}
\label{sec:inc-coplanar}

Our simulations in Section~\ref{sec:fiducial} assume that the initial orbits of the two BHs have inclinations $i_1=i_2=\RH/a_1$ and random longitudes of ascending nodes. 
We expect the CE rates to increase when $i_1$ and $i_2$ are smaller.
Figure~\ref{fig:NCE-vs-t-all-inc} shows the cumulative number of CE2 counts (averaged over 2000 runs) in simulations with different initial values of $i_1$ and $i_2$.
We see that, as expected, $\left<N_2(t)\right>$ increase as $i_1$ and $i_2$ decrease. This increase is highly substantial in the exact coplanar systems ($i_1=i_2$=0), for which we find an average of 56 CE2s at $t=10^5P_1$ and the CE2 rate evolve in time as
\be
\label{eq:N-vs-t-coplanar}
\left< N_2(t) \right> \simeq 0.05 \left(\frac{t}{P_1}\right)^{0.61} \quad \text{(exact coplanar)}.
\ee
This should be compared to Equation~\eqref{eq:N-vs-t-fid} for our fiducial runs (which assume $i_1=i_2=\RH/a_1$).
We notice that a smaller (but non-zero) initial inclination leads to more CE2 at the beginning, but the various $\left<N_2(t)\right>$ curves become roughly parallel after a few ten-thousand orbits, indicating that the rate $d\langle N_2 \rangle/dt$ ``saturates'' to the fiducial value (see below).

\begin{figure}[ht]
    \epsscale{1.0}
    \plotone{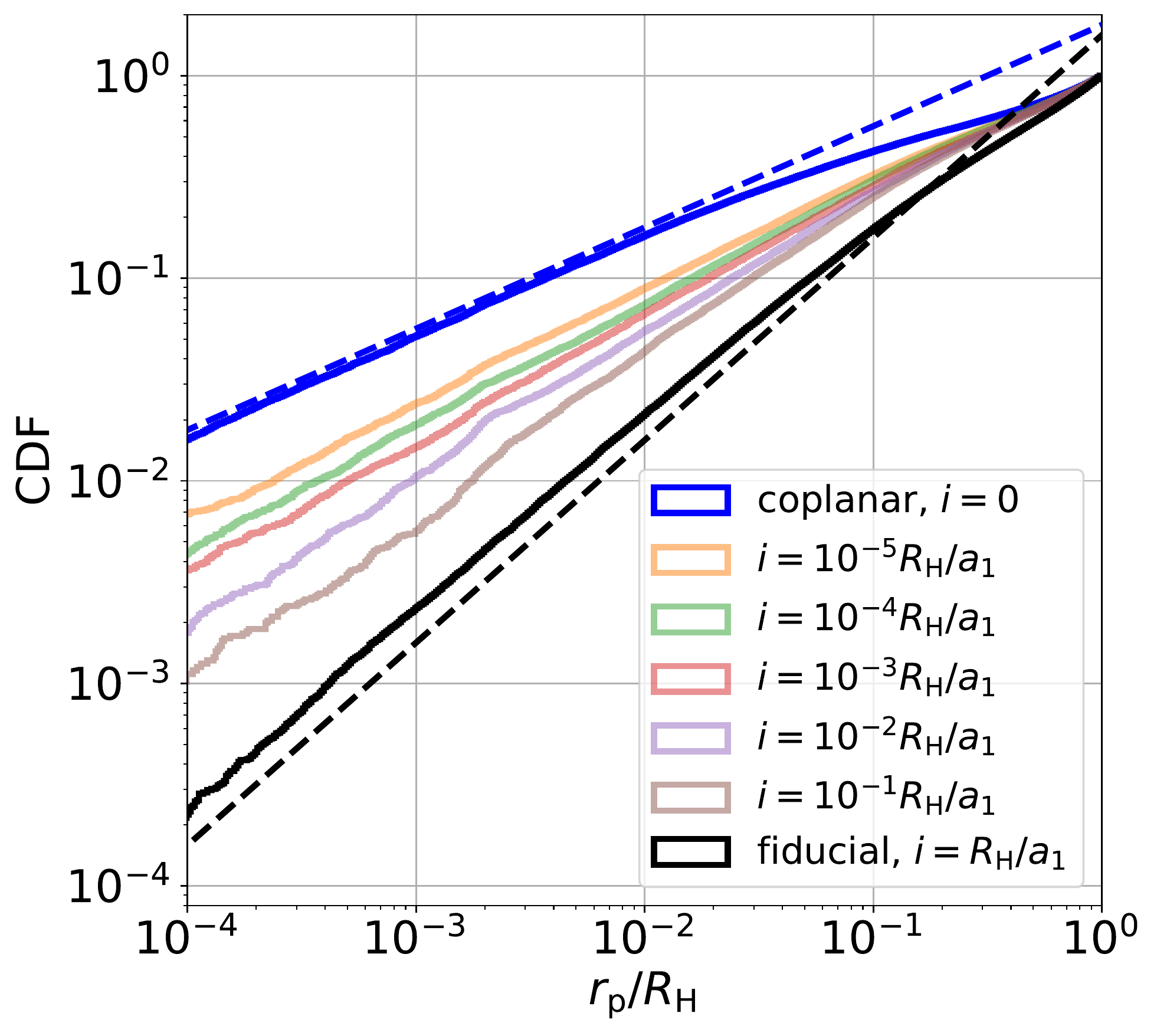}
    \caption{Same as Figure~\ref{fig:CDF-rp}, but showing results for the CE0 events only and for different values of the initial orbital inclinations ($i_1=i_2=i$) as indicated by the labels. The black solid curve is the same as in Figure~\ref{fig:CDF-rp}. The black dashed line shows the linear $\propto \rp$ scaling (see Equation~\ref{eq:Probability-vs-rp}) and the blue dashed line shows the $\sqrt{\rp}$ scaling (see Equation~\ref{eq:Probability-vs-rp-coplanar}).
     }
    \label{fig:CDF-rp-inc}
\end{figure}

The distributions of the pericenter (``closest'') distance $\rp$ of CE2 for the simulations with different initial inclinations are shown in Figure~\ref{fig:CDF-rp-inc}. 
For the exact co-planar systems, we find that the cumulative distribution of $\rp$ is 
\be
\label{eq:Probability-vs-rp-coplanar}
P_{\alpha}(<\rp) \simeq \left(\frac{\rp}{10^{-\alpha}\RH}\right)^{1/2} \quad (\text{for } \rp \leq 10^{-\alpha}\RH).
\ee
This should be contrasted with Equation~\eqref{eq:Probability-vs-rp} for our fiducial simulations.
Equation~\eqref{eq:Probability-vs-rp-coplanar} is equivalent to a uniform distribution of angular momentum, $dP/d\ell_{\rm rel} = \text{const}$. 
Such a uniform distribution is expected because for $i_1=i_2=0$, the dynamics is confined to the original orbital plane, and at $\rrel\ll\RH$ the projected separation $r_{\perp}$ is restricted to a line with $dP/dr_{\perp} = \text{const}$.

Combining Equations~\eqref{eq:N-vs-t-coplanar} and~\eqref{eq:Probability-vs-rp-coplanar}, we find that for the exact co-planar systems, the cumulative BH binary formation rate in very close encounters is 
\be
\label{eq:Nprod-flat}
\langle N(t; \rp<\rcap) \rangle
\simeq 0.005 \left(\frac{t}{P_1}\right)^{0.61} \left(\frac{\rcap}{10^{-4}\RH}\right)^{1/2}.
\ee
It would only take about $6\times10^{3}P_1$ on average for the two BHs to form a merging binary.
Compared to Equation~\eqref{eq:Nprod}, we see that for such perfectly coplanar systems, the BH binary formation rate is much higher than our fiducial systems (with $i_1=i_2=\RH/a_1$).

All of the other small initial inclination simulations yield between 0.1 and 1.0 critical CEs with $\rp<10^{-4}\RH$ on average in their first $10^{5}P_1$. We expect their rates to be similar to the fiducial rates over longer timescales (see below).

\begin{figure}[ht]
    \epsscale{1.0}
    \plotone{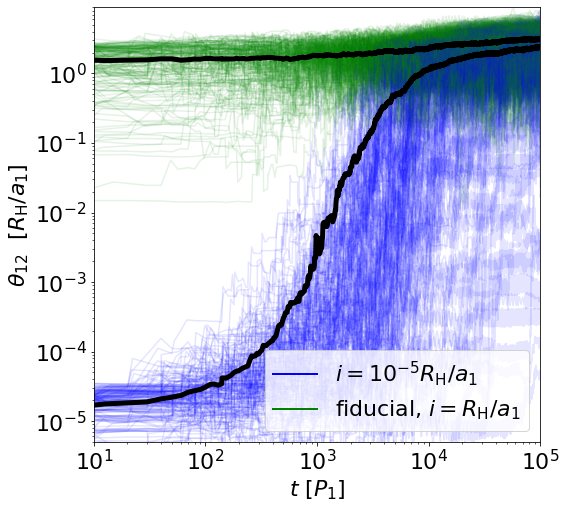}
    \caption{Time evolution of the mutual inclination of the two BH orbits in the fiducial system (green) and in the nearly coplanar system with initial $i_1=i_2=10^{-5}\RH/a_1$ (blue). The colored lines represent 200 individual simulations for each system. The black curves represent their averages.
    }
    \label{fig:theta-vs-t}
\end{figure}

\subsection{Evolution of nearly coplanar systems}
\label{sec:inc-time}

Figures~\ref{fig:NCE-vs-t-all-inc} and~\ref{fig:CDF-rp-inc} show that for nearly coplanar systems (with initial inclinations $0<i_1,i_2\ll \RH/a_1$), both the cumulative CE2 rate and the $\rp$ distribution $P_{\alpha}(<\rp)$ lie between those of the exact coplanar system and our fiducial system.
However, Figure~\ref{fig:NCE-vs-t-all-inc} also indicates that while the cumulative number of CE2 events in the first $\sim 10^4$ orbits is much larger than the fiducial case, the rate $d\left<N_2\right>/dt$ seems to settle down to the fiducial rate at later times. 
This suggests that the mutual inclinations of the BH orbits grow in time in systems with nearly coplanar initial orbits.

Figure~\ref{fig:theta-vs-t} shows the evolution of mutual inclination $\theta_{12}$ of the BH orbits for our fiducial systems and for a system with initial $i_1=i_2=10^{-5}\RH/a_1$. 
The mutual inclination is computed from $\cos{\theta_{12}}=\hat{\ell}_1\cdot\hat{\ell}_2$, where $\hat{\ell}_1$ and $\hat{\ell}_2$ are the unit angular momentum vectors of $m_1$ and $m_2$ around the SMBH.
We see that in the nearly coplanar system, the mutual inclination $\theta_{12}$ gradually increases and then saturates at $\theta_{12}\sim\RH/a_1$. 
For our fiducial system, the average mutual inclination remains at $\sim\RH/a$ throughout the simulation. 

\begin{figure}[ht]
    \epsscale{1.0}
    \plotone{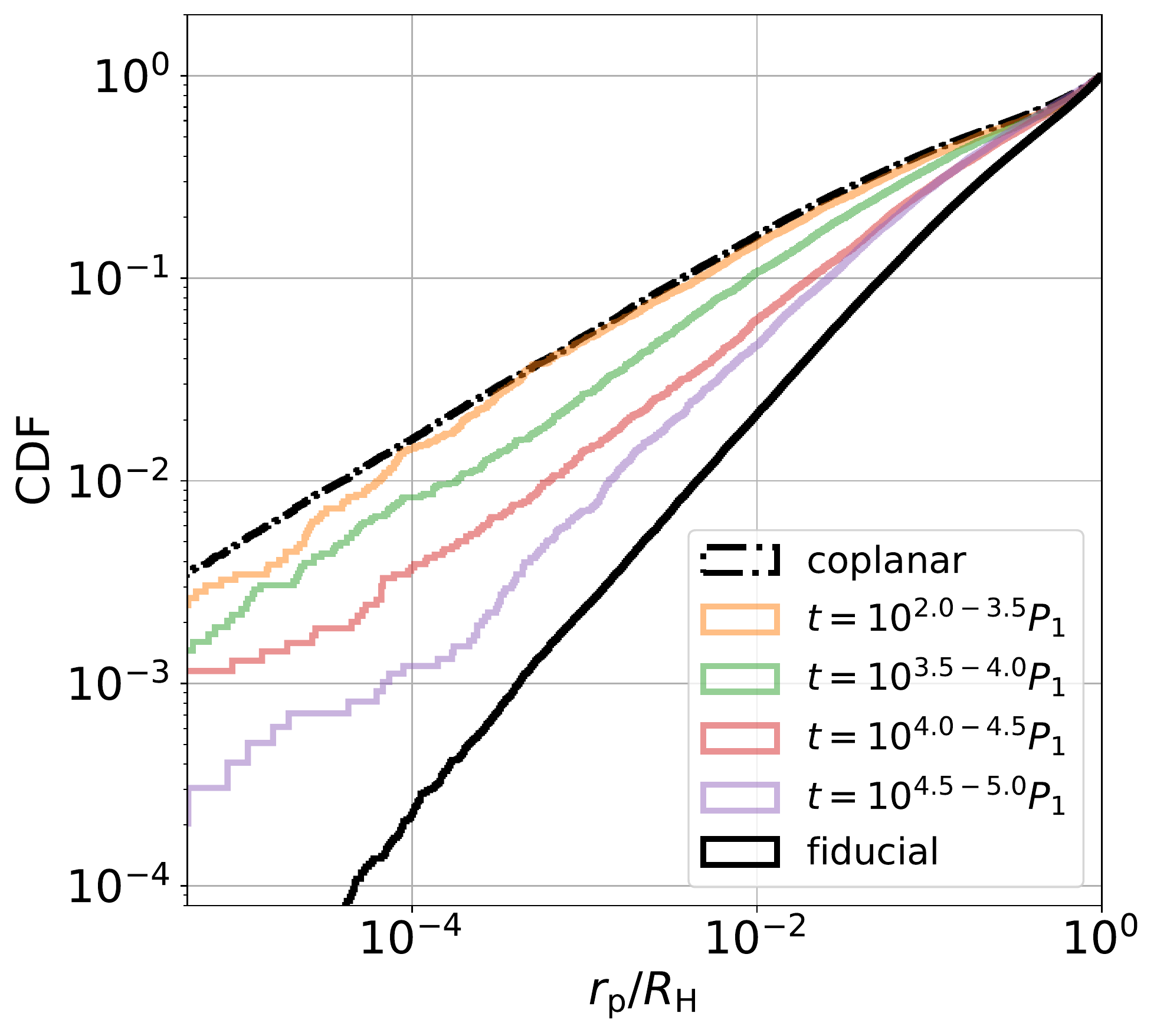}
    \caption{Same as Figure~\ref{fig:CDF-rp-time}, showing the time evolution of the cumulative distribution of $\rp$. The colored lines show the results for a system with initial inclination $i_1=i_2=10^{-5}\RH/a_1$, and different colors indicate different time intervals of the simulations. The black lines show the total distribution for the exact coplanar system (dash-dot) and the fiducial system (solid) for comparison. 
    }
    \label{fig:CDF-rp-inc-vs-time}
\end{figure}

Due to the time evolution of the mutual inclination $\theta_{12}$, we expect that the CE statistics, such as the $\rp$-distribution, also evolve with time in the nearly coplanar systems.
Figure~\ref{fig:CDF-rp-inc-vs-time} shows the time dependence of the $\rp$-distribution for the $i_1=i_2=10^{-5}$ system. 
It is clear that the initial small inclinations ($i_1$, $i_2\ll\RH/a_1$) only affect the results at earlier times. 
The long-term statistics of CEs for nearly coplanar systems is better represented by our fiducial simulations with $i_1=i_2=\RH/a_1$ (see Section~\ref{sec:fiducial}).

\section{Effects of Frictional Disk Forces}
\label{sec:friction}

The AGN disk can affect the dynamical evolution of the embedded BHs. 
For example, a BH (with mass $m_1$ and semi-major axis $a_1$) experiences eccentricity damping on the timescale \citep{Tanaka2004}
\be
\nonumber
\tau_{e} & \simeq & \frac{M^2h^4}{2\pi m_1 \Sigma a_1^2}P_1 \\
\nonumber
& \simeq &
1.2\times10^6 \left(\frac{a_1}{10^{2}GM/c^2}\right)^{-2}\left(\frac{\Sigma}{10^5\text{g/cm}^2}\right)^{-1} \\
& & \times \left(\frac{h}{0.03}\right)^4\left(\frac{m_1}{10M_\odot}\right)^{-1}P_1,
\ee
where $\Sigma$ is the disk surface density and $h$ is the disk aspect ratio, and we have adopted some representative parameters for AGN disks (e.g., \citealp{Sirko2003}; see Figure 1 of \citealp{Secunda2019}). 
In the previous sections, we have studied BH binary captures via very close encounters in the absence of any disk force on the BHs. 
A full exploration of the effects of disk forces on BH binary formation would require long-term hydrodynamical simulations and is beyond the scope of this paper. 
Here, to qualitatively assess the effect of the disk, we apply simple prescriptions of disk forces on the BHs in our $N$-body simulations.

We consider two simple models for the disk forces:
\begin{enumerate}

    \item The first model includes the frictional force (per unit mass):
    \be
    \label{eq:df-drag}
    \vec{F}_{\rm drag} = -\frac{\vec{v} - \vec{v}_{\rm K}}{\tau_{\rm drag}},
    \ee
    where $\vec{v}_{\rm K}=\sqrt{GM/r^3}\ \hat{\theta}$ is the Keplerian velocity and $r$ is the instantaneous distance of the BH to the SMBH. This force tends to damp the BH velocity $\vec{v}$ to $\vec{v}_{\rm K}$ and damp its eccentricity around the SMBH at the rate $\dot{e}\simeq-e/\tau_{\rm drag}$.
    
    \item The second model includes a force that mimics a migration trap at radius $r_0$:
    \be
    \label{eq:df-trap}
    \vec{F}_{\rm trap} = -\frac{\Omega_{\rm K,0} (r-r_0)}{\tau_{\rm trap}}\hat{\vec{\theta}},
    \ee
    where $\Omega_{\rm K,0}=\sqrt{GM/r_0^3}$ is the Keplerian frequency at $r_0$. Equation~\eqref{eq:df-trap} assumes that the torque on the BH at $r$ is approximately linear in $(r-r_0)$ near the trap. In the following,  we set $r_0$ to $a_1$, the initial semi-major axis of $m_1$. 
\end{enumerate}
The constants $\tau_{\rm drag}$ and $\tau_{\rm trap}$ in Equations~\eqref{eq:df-drag} to~\eqref{eq:df-trap} characterize the strengths of the disk forces. We apply these disk forces to our fiducial systems (ses Section~\ref{sec:fiducial}), considering different values of $\tau_{\rm drag}$ and $\tau_{\rm trap}$. For each value case, we perform 200 simulations up to $10^5$ orbit.

\begin{figure}[ht]
    \epsscale{1.2}
    \plotone{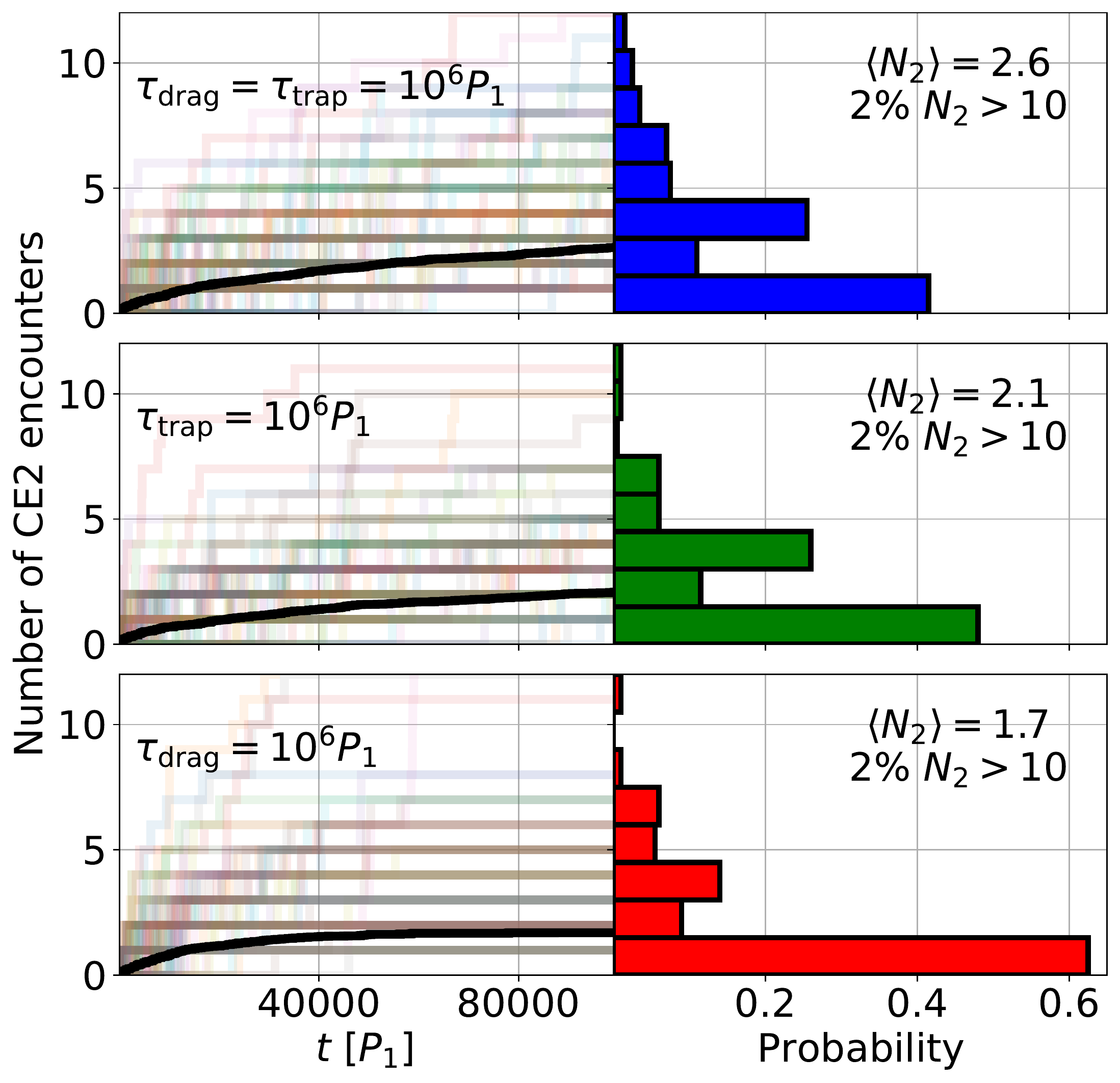}
    \caption{Similar to Figure~\ref{fig:NCE-vs-t-3v}, but for the fiducial systems (see Figure~\ref{fig:NCE-vs-t}) and including the effects of disk forces. 
    The type and the strength (timescale) of the adopted disk force for each panel is as labeled.
    }
    \label{fig:NCE-disk}
\end{figure}

Figure~\ref{fig:NCE-disk} presents three example simulations: $\vec{F}_{\rm drag}$ only with $\tau_{\rm drag}=10^{6}P_1$, $\vec{F}_{\rm trap}$ only with $\tau_{\rm trap}=10^{6}P_1$, and both $\vec{F}_{\rm drag}$ and $\vec{F}_{\rm trap}$ with $\tau_{\rm drag}=\tau_{\rm trap}=10^{6}P_1$.
The left panel shows the time evolution of $N_2$ and the right panel shows distribution of $N_2$ at $t=10^5P_1$.
In all three cases, we have the average $\langle N_2 (t=10^5P_1)\rangle \simeq 2$.
Due to the stochastic nature of the evolution, all three simulations exhibit large variation in the individual $N_2$, with about $2\%$ of the runs having $N_2(t=10^5P_1)\gtrsim10$.

\begin{figure}[ht]
    \epsscale{1.0}
    \plotone{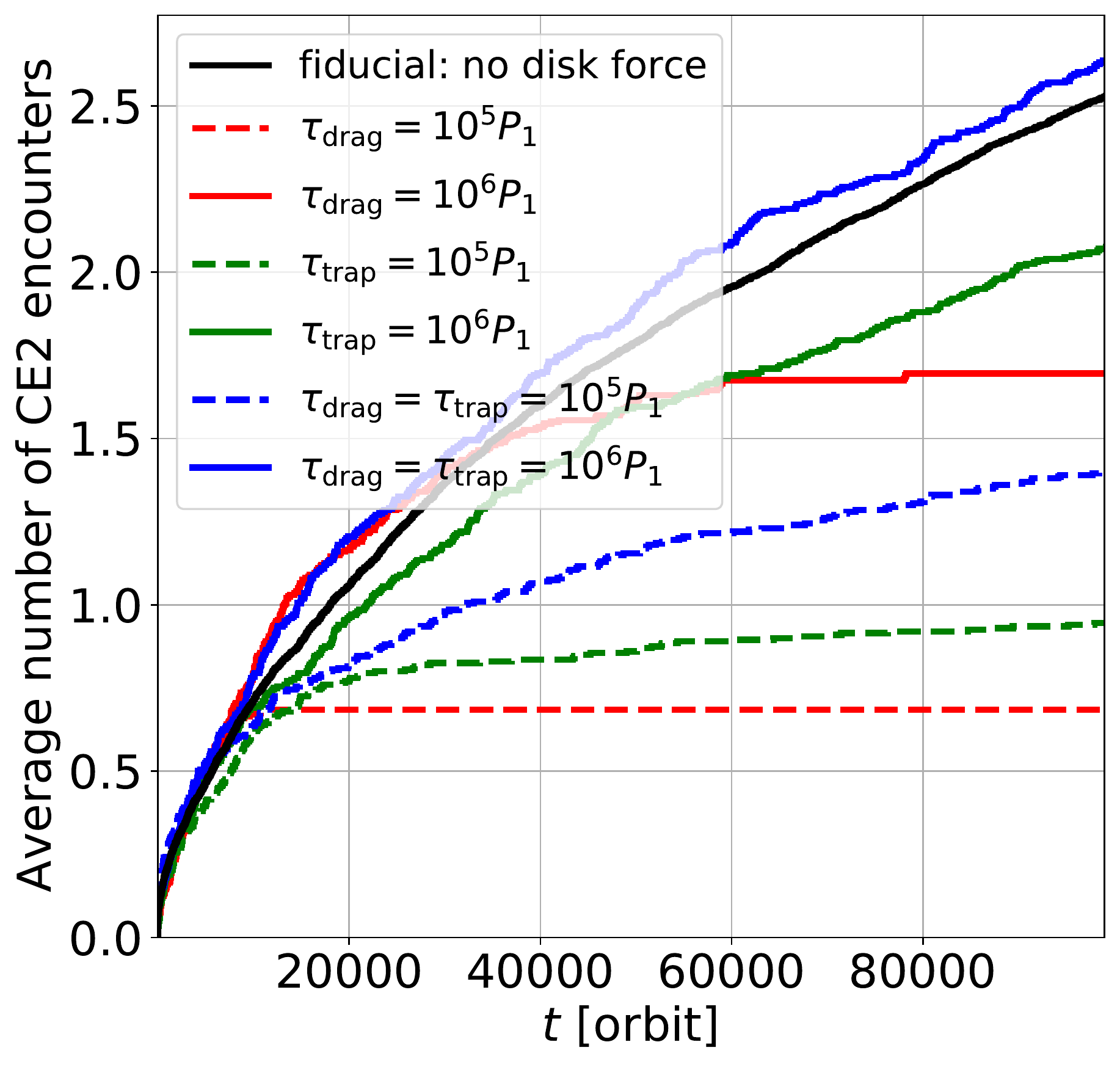}
    \caption{Average number of CE2 events as a function of the time in the fiducial systems (see Fig.~\ref{fig:NCE-vs-t}), including different types and strengths of disk forces. For example, the red solid line refers to the case that includes only $\vec{F}_{\rm drag}$ (with $\tau_{\rm drag}=10^{6}P_1$), the green solid line includes only $\vec{F}_{\rm trap}$ (with $\tau_{\rm trap}=10^6P_1$), and the blue solid line includes both $\vec{F}_{\rm drag}$ and $\vec{F}_{\rm trap}$ (with $\tau_{\rm drag}=\tau_{\rm trap}=10^6P_1$).
    }
    \label{fig:NCE-vs-t-tau}
\end{figure}

\begin{figure}[ht]
    \epsscale{1.2}
    \plotone{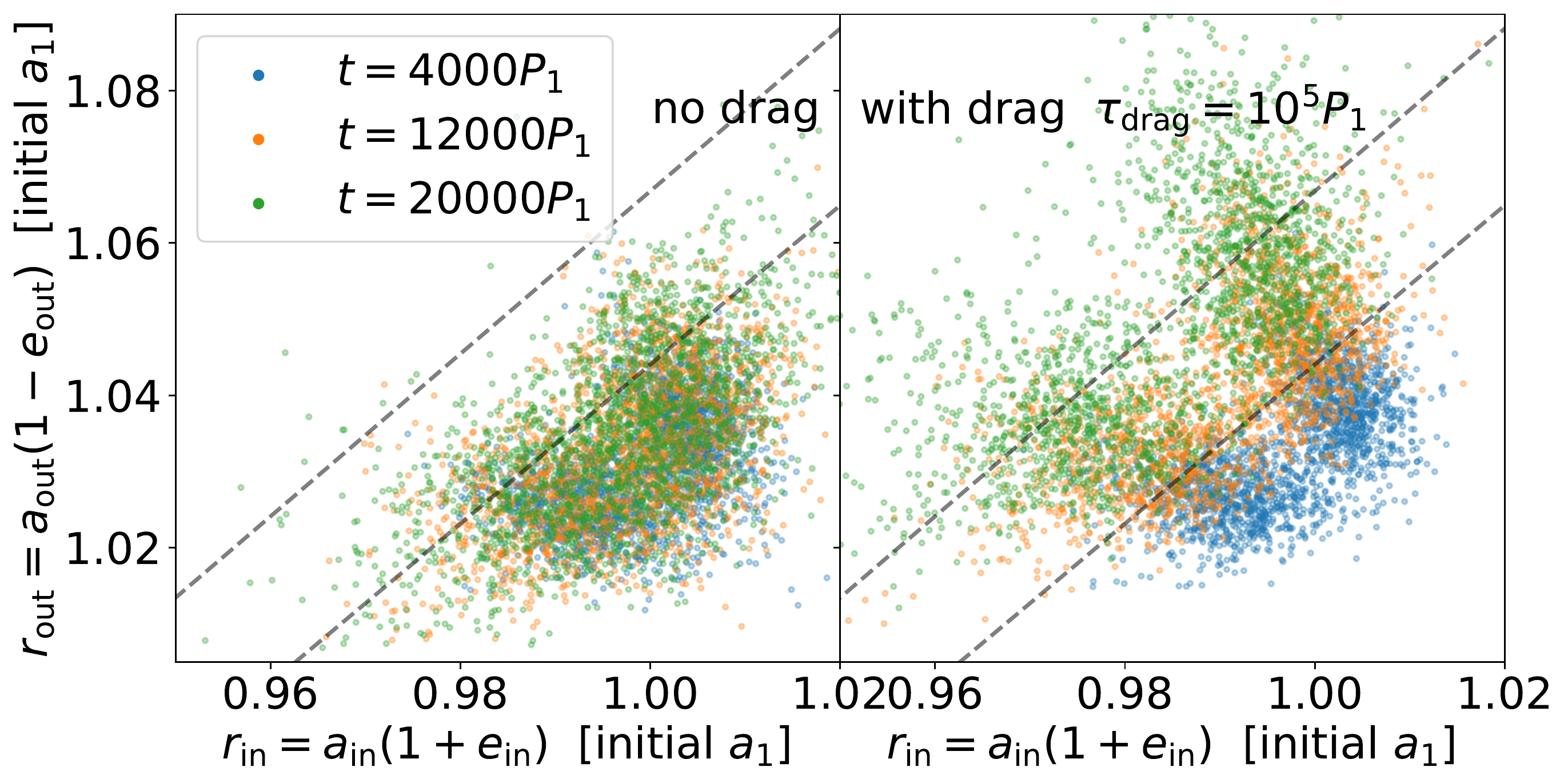}
    \caption{Apocenter of the inner BH $r_{\rm in} = a_{\rm in}(1+e_{\rm in})$ and the pericenter of the outer BH $r_{\rm out} = a_{\rm out}(1-e_{\rm out})$ from the fiducial (left) and the $\tau_{\rm drag}=10^{5}P_1$ simulations (right) at three different times. 
    The region between the dashed lines corresponds to where $(r_{\rm out} - r_{\rm in})/\RHot \in (2,3)$, with $\RHot$ given by Equation~\eqref{eq:RH} and assuming $a_{1,2}\simeq r_{\rm in,out}$. 
    It is marked in both panels to help comparing the distributions. 
    }
    \label{fig:scatter-rin-vs-rout-tau}
\end{figure}

Figure~\ref{fig:NCE-vs-t-tau} shows the effects of different disk forces on the CE2 rates by comparing the time evolution $\langle N_2(t)\rangle$ from our simulations with different force types and strengths.
The drag force (Equation~\ref{eq:df-drag}) tends to stabilize the system by preventing crossing orbits between the BHs.
When $\vec{F}_{\rm drag}$ with $\tau_{\rm drag}=10^5P_1$ (dashed red) or $10^6P_1$ (solid red) is applied, the system is still unstable initially and the two BHs experience CE2 events (also see Li, Rodet and Lai, in prep).
However, no more CE2s are found after about $t=10^4P_1$ for $\tau_{\rm drag}=10^5P_1$ and $t=8\times10^4P_1$ for $\tau_{\rm drag}=10^6P_1$.
To check why CEs cease, Figure~\ref{fig:scatter-rin-vs-rout-tau} compares the orbital separation of the two BHs in the no-drag simulations (fiducial, left panel) and the with-drag simulations ($\tau_{\rm drag}=10^{5}P_1$, right panel) at $t=4\times10^{3}P_1$, $1.2\times10^{4}P_1$, and $2\times10^{4}P_1$. 
In the fiducial simulations, $r_{\rm in}=a_{\rm in}(1+e_{\rm in})$ (the apocenter distance of the inner BH to the SMBH) and $r_{\rm out}=a_{\rm out}(1-e_{\rm out})$ (the pericenter distance of the outer BH) spread to the region with $r_{\rm out}-r_{\rm in} \lesssim 2.5 \RHot$ during the first 4000 orbits and remain in the same region at later time. 
This allows the two BHs to continue to ``engage'' with each other.
However, in the $\tau_{\rm drag}=10^{5}P_1$ simulations, the BH orbits evolve in time toward smaller $r_{\rm in}$ and larger $r_{\rm out}$.
Eventually, the difference between $r_{\rm out}$ and $r_{\rm in}$ becomes too large and CE becomes very rare. 
As a result, no CE2 happens between $t=2\times10^{4}P_1$ and $10^{5}P_1$ in the with-drag simulations. 

One may expect that the trapping force (Equation~\ref{eq:df-trap}) can accelerate the CEs by keeping the BH orbits close to the trapping radius.  
Our simulations show that when $\vec{F}_{\rm trap}$ is applied, our systems indeed have more CE0 events.
However, Figure~\ref{fig:NCE-vs-t-tau} shows that the CE2 rate is actually smaller than the fiducial simulations without $\vec{F}_{\rm trap}$:
The average number of CE2 events becomes $2.1$ for $\tau_{\rm trap}=10^6P_1$ (solid green) and $0.9$ for $\tau_{\rm trap}=10^5P_1$ (dashed green) after $10^5$ orbits. 
Unlike in the simulations with $\vec{F}_{\rm drag}\neq0$, the CE2 rate in our $\vec{F}_{\rm trap}\neq0$ simulations do not drop to zero at later times.

The stabilizing effects of the drag force and the trapping force can balance each other.
The blue curves in Figure~\ref{fig:NCE-vs-t-tau} show that applying $\tau_{\rm drag}=\tau_{\rm trap}=10^{5}P_1$ produces more CE2 events than applying either of the two forces. With $\tau_{\rm drag}=\tau_{\rm trap}=10^{6}P_1$, the two BHs experience slightly more CE2 events than in the fiducial simulations (black curve).

\begin{figure}[ht]
    \epsscale{1.0}
    \plotone{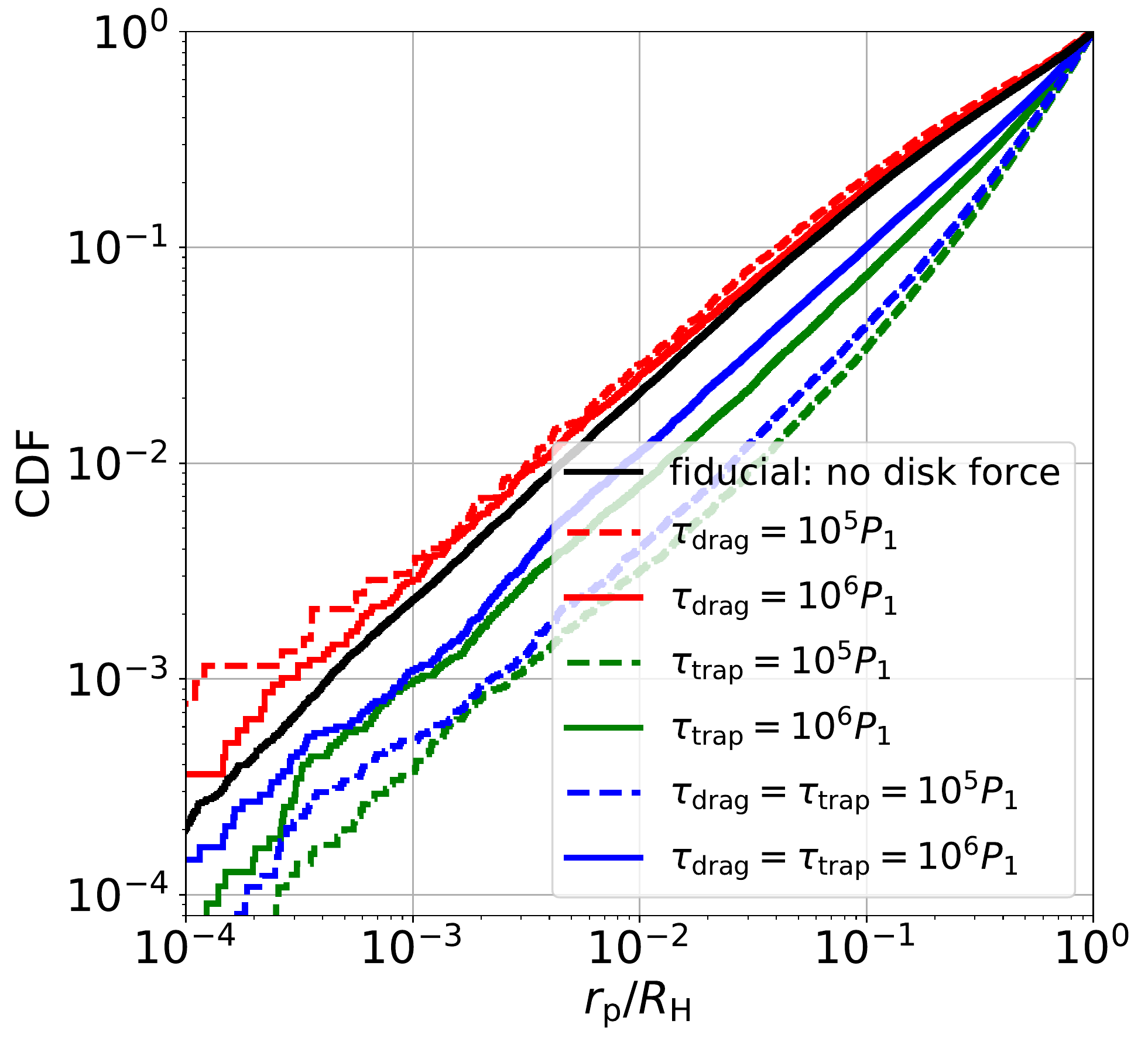}
    \caption{Similar to Figure~\ref{fig:CDF-rp}, but showing the cumulative distribution of $\rp$ in CE0 events for systems with different types and strengths of disk forces.
    }
    \label{fig:CDF-rp-tau}
\end{figure}

Figure~\ref{fig:CDF-rp-tau} shows that the $\rp$-distributions of our simulations with drag forces are similar to the fiducial result. 
In particular, for $\rp/\RH\lesssim 0.01$, the linear scaling of the distribution with $\rp$ remains accurate in all cases. 
In the simulations with the trap force, the mild encounters with $\rp/\RH\gtrsim0.01$ are relatively more frequent than in the fiducial system with no disk forces (as the CDFs with the trap force have a larger slope at $\rp/\RH\gtrsim0.01$). 
Thus the simple power-low distribution of $\rp$ given by Equation~\eqref{eq:Probability-vs-rp} is still valid for CE2 but becomes less accurate for CE0 when $\vec{F}_{\rm trap}$ is applied.

In a real AGN disk, multiple effects can take place at the same time. 
The outcome of their competition depends on the detailed disk properties. 
Our results in this section suggest that disk forces have little effect on the $\rp$-distribution for very close encounters (Figure~\ref{fig:CDF-rp-tau}), but can influence the CE2 event rate (Figure~\ref{fig:NCE-vs-t-tau}), and therefore affecting the BBH formation rate.
Our simulations also suggest that in order for the two unstable BHs in a AGN disk to capture into a merging binary, a combination of the drag force and trapping force from the disk is needed, or we would rely on the chance that an individual system is in the high-$N_2$ tail of the distribution.
We emphasize that our results in this section are based on simple prescriptions of disk forces. 
Hydrodynamics simulations will be needed to fully capture the effects of disk forces.

\section{Three and more BH\lowercase{s} around SMBH}
\label{sec:NBH}
While a ``SMBH + 2BHs'' system is unstable only if $a_2-a_1 \lesssim 2\sqrt{3}\RH$ (see Equation~\ref{eq:a1-a2-criterion}), no precise stability criterion exists for systems with more than two BHs.
In fact, such systems always exhibit instability eventually, except that the instability time grows with increasing orbital spacings.
Numerical integrations suggest that a system of 3 or more bodies on nearly circular, coplanar orbits around a central massive object is stable for at least $N$ orbital periods if the separation between adjacent bodies satisfies $\left|a_{j+1}-a_{j}\right|\gtrsim K(N)\RH$, where the constant $K(N)$ increases with increasing $N$ \citep[e.g., $K\simeq9-12$ for $N=10^{10}$;][]{Smith2009}.

To explore close encounters in systems with more than two BHs, we consider a ``SMBH + 3BHs'' system with $(m_1,m_2,m_3)=(2,1,0.5)\times10^{-5}M$ and a ``SMBH + 5 BHs'' system with $(m_1,m_2,m_3,m_4,m_5)=(2,1,0.5,0.25,0.125)\times10^{-5}M$, where $M$ is mass of the central SMBH. The initial orbital separation is set as
\be
a_{j+1}-a_j = K \RHm,
\ee
where $\RHm$ is mutual Hill radius of $m_j$ and $m_{j+1}$ (see Equation~\ref{eq:RH}) and we set $K=2$ or $4$.
The BHs are given initial eccentricities $e_1=0$, $e_j=10^{-5}$ for $j>1$ and inclinations $i=R_{\rm H}/a_1$ for all BHs. 
Similar to our fiducial simulations (Section~\ref{sec:fiducial}), we carry out 2000 runs and sample the initial values of the argument of the peripasis, the longitude of the ascending node, and the mean anomaly randomly in the range $[0,2\pi]$ for each BH, assuming they all have uniform distributions.

\begin{figure}[ht]
    \epsscale{0.9}
    \plotone{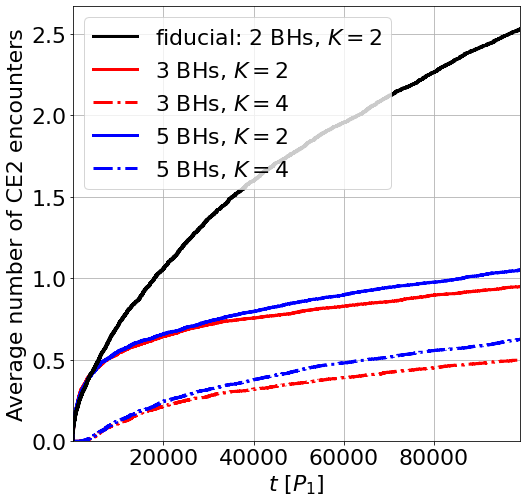}
    \caption{Average number of CE2 as a function of time for systems with 3 and 5 BHs. The fiducial result for 2 BHs (the bottom left planel of Figure~\ref{fig:NCE-vs-t}) is also shown for comparison. 
    }
    \label{fig:NCE-vs-t-3p}
\end{figure}

Figure~\ref{fig:NCE-vs-t-3p} shows the averaged cumulative CE2 rates for systems with two, three and five BHs. 
The two-BH (fiducial) systems experience nearly 3 to 5 times more CE2 events than the other systems.
For the $K=2$ cases, the average number of CE2 events at $t=10^5P_1$ is 0.9 with 3 BHs and 1.1 with 5 BHs.
The fact that the CE2 rate is highest for the two-BH systems is somewhat surprising.
It may be because the two-BH systems are subjected to conservation constraints, which cause the BH orbits to repeatedly overlap. 
A three-or-more-BH architecture has more degrees of freedom and there is no conservation law between any two BHs.

Adopting $K=4$ in the three- and five-BH simulations leads to no CE2 events before $t \simeq 2\times10^{3}P_1$. 
This is because the initial conditions with larger $K$ require longer times for the BHs to develop their first Hill sphere crossing. 
After the instability develops, the $K=4$ systems catch up in the growth rate. 
At $t=10^5P_1$, the three-BH and five-BH simulations with $K=4$ yield $\left< N_2 \right>=0.5$ and $0.62$, respectively. 
For systems with larger $K$, we expect a longer time before the CEs occur, but the CE rate will eventually converge.

We fit the time evolution of the average number of CE2 events for $t>5\times10^4P_1$, by
\be
\left< N_2(t) \right> = \left(\frac{t}{T}\right)^{n_2}.
\ee
Least-square fit gives $n_2 = 0.26$, $0.50$, $0.30$, $0.50$ and $T = 1.2\times10^5$, $4.0\times10^5$, $8.5\times10^4$, $2.6\times10^5$  for 3 BHs with $K=2$, 3 BHs with $K=4$, 5 BHs with $K=2$ and 5 BHs with $K=4$, respectively. 
Using $\langle N(t;\rp<\rcap) \rangle = \langle N_2 (t) \rangle P_2(<\rcap)$ (see Equation~\ref{eq:N-rcap-t}) with $P_2(<\rcap) \simeq \rp/ (10^{-2}\RH)$ (see Equation~\ref{eq:Probability-vs-rp}), we find that it would take about $5\times10^{12}P_1$, $4\times10^{9}P_1$, $4\times10^{11}P_1$, $3\times10^{9}P_1$ on average to form a merging BH binary in each of the four cases.

\begin{figure}[ht]
    \epsscale{1.0}
    \plotone{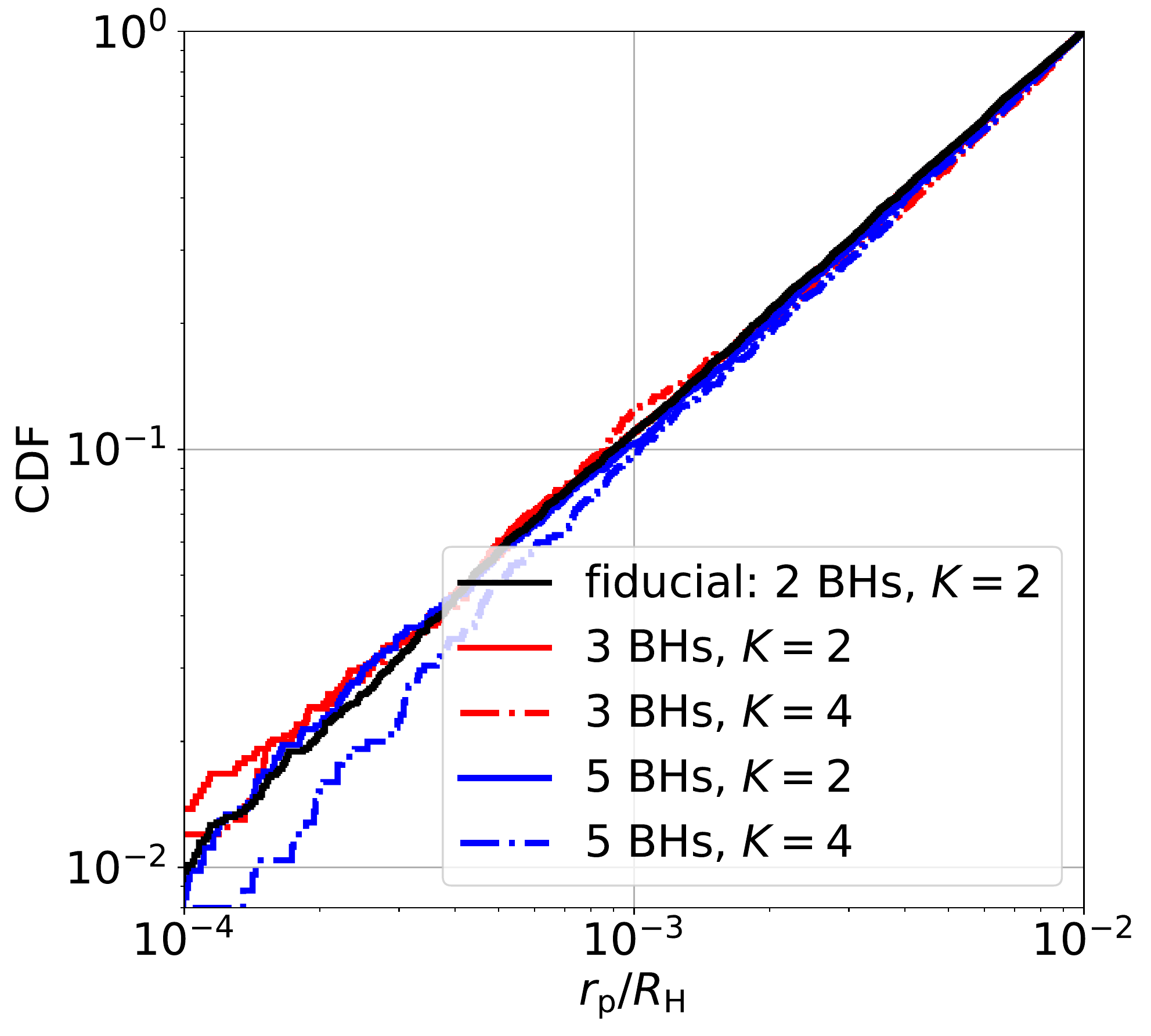}
    \caption{Same as Figure~\ref{fig:CDF-rp}, except showing the cumulative distribution of $\rp$ in close encounters (CE2 only) in simulations with different number of BHs and initial spacings. 
    }
    \label{fig:CDF-rp-3p}
\end{figure}

\begin{figure}[ht]
    \epsscale{1.0}
    \plotone{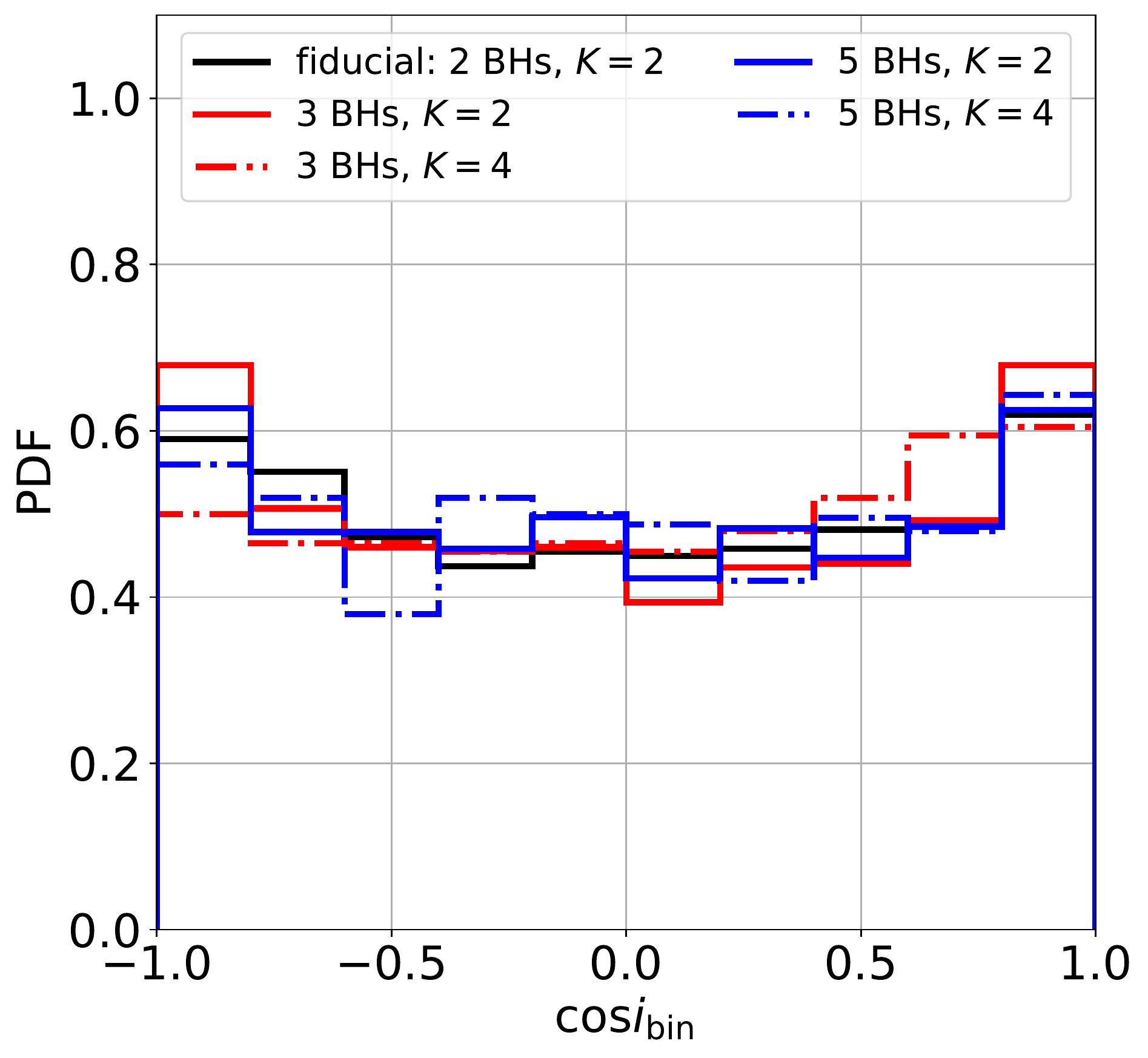}
    \caption{Same as Figure~\ref{fig:PDF-inc}, except showing the results for the CE2 events in the simulations with different number of BHs and initial spacings.
    }
    \label{fig:PDF-inc-3p}
\end{figure}

Figure~\ref{fig:CDF-rp-3p} shows that the $\rp$-distributions of CE2 for all of the simulations in this section are similar to the fiducial result (see Figure~\ref{fig:CDF-rp}).
Figure~\ref{fig:PDF-inc-3p} shows that CE2 always have nearly uniform distributions of $\cos{i_{\rm bin}}$ and thus the merging BH binaries formed in our scenario have a wide range of inclination angles with respect to the AGN disk plane. 
This is expected: Regardless of the number and the initial spacings of the BHs, the relative orbit of the captured binary is dominated by the gravity of the BHs during the CEs. 
Having more BHs and different initial spacings only affects the CE rate.

\section{Summary and Discussion}
\label{sec:summary}

In this paper, we have studied the long-term evolution of two or more
stellar-mass BHs on closely-packed, dynamically unstable circular orbits
around a SMBH, with the initial semi-major axes separated by a few
times the Hill radius $\RH$.  Such an orbital configuration may
naturally arise from BH migrations in the AGN disk and leads to
recurring close encounters between the BHs. We use $N$-body
simulations to study the statistics and rate of close encounters (of
various degrees of ``closeness'' compared to $\RH$), the properties
of the relative orbits of the encountering BHs, and the probability
for two BHs to be captured into a merging binary with the
help of gravitational wave emission during very close encounters. Our
fiducial simulations focus on ``SMBH + 2 BHs'' systems with
$m_1=2m_2=2\times10^{-5}M$ and initial inclinations $i_1=i_2=\RH/a_1$
(where $a_1$ and $a_2$ are the BHs' semi-major axes around the SMBH).
Additions simulations with different BH masses, initial orbital
inclinations, prescriptions for disk forces, and different
number of BHs are also performed.

Our simulations show that close encounters (CEs) between the BHs in such
systems exhibit three general characteristics:
\begin{enumerate}
    \setcounter{enumi}{0}
\item 
Close encounters (with the separation between two BHs less than $\RH$)
occur stocastically and the rate of CEs generally declines in time
(see Fig.~\ref{fig:NCE-vs-t}). The average cumulative number of CEs,
$\langle N(t)\rangle$, is approximately a power-law function of time,
although there are wide spreads in $N(t)$ for different systems.  For
our fiducial ``SMBH + 2 BHs'' setup, we find $\langle
N(t)\rangle\propto t^{0.5}$ (see Eqs.~\ref{eq:N-vs-t}-\ref{eq:N-vs-t-fid}).
\item 
The vast majority of the CEs result in the formation of short-lived binaries
with a small binding energy (of order $Gm_1m_2/\RH$; see Fig.~\ref{fig:PDF-aCE}).
Such binaries are quickly destroyed by the SMBH tide and do not lead to binary mergers.
\item 
The closest separation ($\rp$) during a close encounter follows a
cumulative distribution $P(<\rp)\propto\rp$ for $\rp\ll\RH$ (see
Figs.~\ref{fig:CDF-rp},~\ref{fig:CDF-rp-time}, and Eq.~\ref{eq:Probability-vs-rp}).  This distribution is robust,
regardless of the BH masses (Section~\ref{sec:fid-mass}) and the number of BHs in the system
(Section~\ref{sec:NBH}). For systems with very small but non-zero initial mutual inclinations,
the same $\rp$-distribution applies at later times as the mutual inclinations grow.
In a system with two exact co-planar BHs, the distribution becomes
$P(<\rp)\propto\rp^{1/2}$ (Fig.~\ref{fig:CDF-rp-inc} and Eq.~\ref{eq:Probability-vs-rp-coplanar}).
\end{enumerate}
These results imply that, to capture an encountering BH pairs into a
``permenant'' binary, a fast dissipative mechanism is required.
Given the high close encounter rate, a promising 
mechanism is the GW emission when $\rp$ is very small
(less than the critical ``capture'' radius $r_{\rm cap}\sim 10^{-4}\RH$,
depending on the system parameters; see Eq.~\ref{eq:GWCE-rp_crit}):
\begin{enumerate}
    \setcounter{enumi}{3}
\item 
We provide a semi-analytical formula for the averaged cumulative rate
of binary captures via GW emission in a ``SMBH + 2 BHs'' fiducial system,
Equation~\eqref{eq:N-rcap-t} or~\eqref{eq:Nprod}, which is the product of the close encounter rate $\langle N(t)\rangle$
and the capture probability $P(<\rcap)$ during each encounter.
\item 
Our formula suggests that, in the fiducial systems, the timescale for
two BHs to be captured is $10^8P_1$ on average (assuming $r_{\rm cap}=10^{-4}\RH$)
and $\lesssim 10^7P_1$ for the $5\%$ of the systems with $N(t)\gtrsim4\langle N(t)\rangle$.
In the exact co-planar systems, the capture rate is much higher (see Eq.~\ref{eq:Nprod-flat}), and
we find that the average binary capture time is $6\times10^3P_1$.
\item
After the two BHs are captured, we expect these BH binaries to merge in a few
binary orbits. Their mergers will exhibit high eccentricities (see below) when
entering the LIGO band ($\gtrsim 10$~Hz) and will have a board distribution of orbital
inclinations relative to the original AGN disk (see Figure~\ref{fig:PDF-inc}).
\end{enumerate}

We have carried out additional simulations to assess how the above results
may be influenced by various system parameters and by the gas disk effects:
\begin{enumerate}
    \setcounter{enumi}{6}
\item 
The masses of the BHs (relative to the mass of the SMBH) affect the close encounter rate
only in a modest way (Section~\ref{sec:fid-mass}). For example, the rate of CEs with $\rp\le 10^{-2}\RH$ 
decreases by a factor of $\sim 2$ when the BH masses are lowered by a factor of ten
(see Figure~\ref{fig:NCE-vs-t-3v}). The $\rp$-distribution (Eq.~\ref{eq:Probability-vs-rp}) is robust. Thus,
we expect that our fiducial binary capture rate, Equation~\eqref{eq:Nprod}, remain valid for systems with
different BH masses.
\item 
The most optimal setup to get more binary captures in our
scenario is with two exact co-planar BHs (see points 3 and 5 above).
Such an exact co-planar system would have a higher rate of close encounters
(see Fig.~\ref{fig:NCE-vs-t-all-inc} and Eq.~\ref{eq:N-vs-t-coplanar}) and a flatter $\rp$-distribution (Fig.~\ref{fig:CDF-rp-inc} and Eq.~\ref{eq:Probability-vs-rp-coplanar}),
leading to a greatly enhanced binary capture rate (Eq.~\ref{eq:Nprod-flat}).
However, if the mutual inclination between the BH orbits 
is initially small but non-zero, it will inevitably grow 
and evolve to an ``equilibrium'' value $\sim \RH/a_1$
(see Fig.~\ref{fig:theta-vs-t}), causing $P(<\rp)$ to
converge to the fiducial result (see Fig.~\ref{fig:CDF-rp-inc-vs-time}).
\item
We have explored the effects of gas disks by applying simple
prescriptions of disk forces (see Eqs.~\ref{eq:df-drag} and~\ref{eq:df-trap}) on the BHs in our
N-body simulations.  We find that such prescribed disk forces
do not necessarily lead to an enhanced binary capture rate.
In fact, simple gas drags on the BHs may stop close encounters at late times.
A ``migration trap'' force can sometimes balance the drag force and
maintain the close encounter rate (see Fig.~\ref{fig:NCE-vs-t-tau}).
\item
The number of BHs and their initial spacings in a closely-packed system
do not affect the $\rp$-distribution (Fig.~\ref{fig:CDF-rp-3p}) and the orbital obliquity
distribution (Fig.~~\ref{fig:PDF-inc-3p}) during the very close encounters.
We find that systems with more than 2 BHs have a lower close encounter rate
than systems with two BHs.
\end{enumerate}

Regarding point 6 above: Using Equation~\eqref{eq:GWCE-rp_crit} for $r_{\rm cap}$, we find that the
GW emitted by the BH binary at capture has a frequency
\begin{eqnarray}
&& f_{\rm cap}={1 \over \pi}\left({Gm_{12}\over r_{\rm cap}^3}\right)^{1/2}\nonumber\\
&&\qquad =(1.4\,{\rm Hz})\,\eta^{3/7}\left({4\mu\over m_{12}}\right)^{\!-3/7}
  \left({a_{12}\over 100 GM/c^2}\right)^{\!-3/7}\nonumber\\
&&\qquad \times \left({M\over 10^8M_\odot}\right)^{\!-2/7}\left({m_{12}\over 100M_\odot}
  \right)^{\!-5/7}.
\end{eqnarray}
For the adopted parameter values, this frequency lies slightly below
the LIGO band.  Such a newly captured binary will merge within a few
binary orbits.  Using the standard gravitational radiation formulae
\citep{Peters1964}, it is easy to see that when the frequency of the GWs
from the binary enters the LIGO band (>10~Hz), the binary can retain
a very significant ($\gtrsim 0.5$) eccentricity.

The event rate of binary BH mergers produced in our scenario depends
on the population of BHs in AGN disks, which is very
uncertain. Nevertheless, given the binary capture timescale obtained
in this paper, a non-negligible event rate for such BH
mergers may be expected. Perhaps the recently claimed eccentric merger event
GW190521 \citep{Gayathri2022} is an example.

The most uncertain aspects of the present study concern the effects of
gas disks. Our conclusion on the gas effects (point 9 above) should be
considered tentative. For future work, hydrodynamics simulations
should be used for a more in-depth study of the disk effects.
Although we have only considered GW emission in this paper, the
broad range of $\rp$ in close encounters may allow the physical processes at
different distance scales to operate.  For example, interaction between the circum-BH disks
may generate dissipation for the BH pairs during close encounters.
Our results of the close encounter rate and the $\rp$-distribution can
be applied to these alternative dissipation mechanisms to
explore different possibilities of binary BH formation in AGN disks.

\acknowledgments
This work is supported in by NSF grant AST-2107796 and the NASA grant 80NSSC19K0444.


\software{Rebound \citep{Rein2012},  
          Matplotlib \citep{Hunter2007}, 
          NumPy \citep{Walt2011},
          SciPy \citep{Virtanen2020}
          }


\vspace{2cm}


\begin{thebibliography}{}
\expandafter\ifx\csname natexlab\endcsname\relax\def\natexlab#1{#1}\fi
\providecommand{\url}[1]{\href{#1}{#1}}
\providecommand{\dodoi}[1]{doi:~\href{http://doi.org/#1}{\nolinkurl{#1}}}
\providecommand{\doeprint}[1]{\href{http://ascl.net/#1}{\nolinkurl{http://ascl.net/#1}}}
\providecommand{\doarXiv}[1]{\href{https://arxiv.org/abs/#1}{\nolinkurl{https://arxiv.org/abs/#1}}}

\bibitem[{{Abbott} {et~al.}(2019){Abbott}, {Abbott}, {Abbott}, {Abraham},
  {Acernese}, {Ackley}, {Adams}, {Adhikari}, {Adya}, {Affeldt}, \&
  et~al.}]{LIGO2019}
{Abbott}, B.~P., {Abbott}, R., {Abbott}, T.~D., {et~al.} 2019,
  \href{http://dx.doi.org/10.3847/2041-8213/ab3800}{\apjl},
  \href{https://ui.adsabs.harvard.edu/abs/2019ApJ...882L..24A}{882, L24}

\bibitem[{{Abbott} {et~al.}(2021){Abbott}, {Abbott}, {Abraham}, {Acernese},
  {Ackley}, {Adams}, {Adams}, {Adhikari}, {Adya}, {Affeldt}, \&
  et~al.}]{LIGO2021a}
{Abbott}, R., {Abbott}, T.~D., {Abraham}, S., {et~al.} 2021,
  \href{http://dx.doi.org/10.3847/2041-8213/abe949}{\apjl},
  \href{https://ui.adsabs.harvard.edu/abs/2021ApJ...913L...7A}{913, L7}

\bibitem[{{Antonini} \& {Perets}(2012)}]{Antonini2012}
{Antonini}, F., \& {Perets}, H.~B. 2012,
  \href{http://dx.doi.org/10.1088/0004-637X/757/1/27}{\apj},
  \href{https://ui.adsabs.harvard.edu/abs/2012ApJ...757...27A}{757, 27}

\bibitem[{{Banerjee} {et~al.}(2010){Banerjee}, {Baumgardt}, \&
  {Kroupa}}]{Banerjee2010}
{Banerjee}, S., {Baumgardt}, H., \& {Kroupa}, P. 2010,
  \href{http://dx.doi.org/10.1111/j.1365-2966.2009.15880.x}{\mnras},
  \href{https://ui.adsabs.harvard.edu/abs/2010MNRAS.402..371B}{402, 371}

\bibitem[{{Bartos} {et~al.}(2017){Bartos}, {Kocsis}, {Haiman}, \&
  {M{\'a}rka}}]{Bartos2017}
{Bartos}, I., {Kocsis}, B., {Haiman}, Z., \& {M{\'a}rka}, S. 2017,
  \href{http://dx.doi.org/10.3847/1538-4357/835/2/165}{\apj},
  \href{https://ui.adsabs.harvard.edu/abs/2017ApJ...835..165B}{835, 165}

\bibitem[{{Baruteau} {et~al.}(2011){Baruteau}, {Cuadra}, \&
  {Lin}}]{Baruteau2011}
{Baruteau}, C., {Cuadra}, J., \& {Lin}, D.~N.~C. 2011,
  \href{http://dx.doi.org/10.1088/0004-637X/726/1/28}{\apj},
  \href{https://ui.adsabs.harvard.edu/abs/2011ApJ...726...28B}{726, 28}

\bibitem[{{Belczynski} {et~al.}(2010){Belczynski}, {Bulik}, {Fryer}, {Ruiter},
  {Valsecchi}, {Vink}, \& {Hurley}}]{Belczynski2010}
{Belczynski}, K., {Bulik}, T., {Fryer}, C.~L., {et~al.} 2010,
  \href{http://dx.doi.org/10.1088/0004-637X/714/2/1217}{\apj},
  \href{https://ui.adsabs.harvard.edu/abs/2010ApJ...714.1217B}{714, 1217}

\bibitem[{{Belczynski} {et~al.}(2016){Belczynski}, {Holz}, {Bulik}, \&
  {O'Shaughnessy}}]{Belczynski2016}
{Belczynski}, K., {Holz}, D.~E., {Bulik}, T., \& {O'Shaughnessy}, R. 2016,
  \href{http://dx.doi.org/10.1038/nature18322}{\nat},
  \href{https://ui.adsabs.harvard.edu/abs/2016Natur.534..512B}{534, 512}

\bibitem[{{Bellovary} {et~al.}(2016){Bellovary}, {Mac Low}, {McKernan}, \&
  {Ford}}]{Bellovary2016}
{Bellovary}, J.~M., {Mac Low}, M.-M., {McKernan}, B., \& {Ford}, K.~E.~S. 2016,
  \href{http://dx.doi.org/10.3847/2041-8205/819/2/L17}{\apjl},
  \href{https://ui.adsabs.harvard.edu/abs/2016ApJ...819L..17B}{819, L17}

\bibitem[{{Downing} {et~al.}(2010){Downing}, {Benacquista}, {Giersz}, \&
  {Spurzem}}]{Downing2010}
{Downing}, J.~M.~B., {Benacquista}, M.~J., {Giersz}, M., \& {Spurzem}, R. 2010,
  \href{http://dx.doi.org/10.1111/j.1365-2966.2010.17040.x}{\mnras},
  \href{https://ui.adsabs.harvard.edu/abs/2010MNRAS.407.1946D}{407, 1946}

\bibitem[{{Ford} \& {McKernan}(2021)}]{Ford2021}
{Ford}, K.~E.~S., \& {McKernan}, B. 2021,
  \href{https://ui.adsabs.harvard.edu/abs/2021arXiv210903212F}{arXiv e-prints,
  arXiv:2109.03212}

\bibitem[{{Fragione} \& {Kocsis}(2019)}]{Fragione2019}
{Fragione}, G., \& {Kocsis}, B. 2019,
  \href{http://dx.doi.org/10.1093/mnras/stz1175}{\mnras},
  \href{https://ui.adsabs.harvard.edu/abs/2019MNRAS.486.4781F}{486, 4781}

\bibitem[{{Gayathri} {et~al.}(2022){Gayathri}, {Healy}, {Lange}, {O'Brien},
  {Szczepa{\'n}czyk}, {Bartos}, {Campanelli}, {Klimenko}, {Lousto}, \&
  {O'Shaughnessy}}]{Gayathri2022}
{Gayathri}, V., {Healy}, J., {Lange}, J., {et~al.} 2022,
  \href{http://dx.doi.org/10.1038/s41550-021-01568-w}{Nature Astronomy}

\bibitem[{{Gladman}(1993)}]{Gladman1993}
{Gladman}, B. 1993, \href{http://dx.doi.org/10.1006/icar.1993.1169}{\icarus},
  \href{https://ui.adsabs.harvard.edu/abs/1993Icar..106..247G}{106, 247}

\bibitem[{{Graham} {et~al.}(2020){Graham}, {Ford}, {McKernan}, {Ross}, {Stern},
  {Burdge}, {Coughlin}, {Djorgovski}, {Drake}, {Duev}, {Kasliwal}, {Mahabal},
  {van Velzen}, {Belecki}, {Bellm}, {Burruss}, {Cenko}, {Cunningham}, {Helou},
  {Kulkarni}, {Masci}, {Prince}, {Reiley}, {Rodriguez}, {Rusholme}, {Smith}, \&
  {Soumagnac}}]{Graham2020}
{Graham}, M.~J., {Ford}, K.~E.~S., {McKernan}, B., {et~al.} 2020,
  \href{http://dx.doi.org/10.1103/PhysRevLett.124.251102}{\prl},
  \href{https://ui.adsabs.harvard.edu/abs/2020PhRvL.124y1102G}{124, 251102}

\bibitem[{{Gr{\"o}bner} {et~al.}(2020){Gr{\"o}bner}, {Ishibashi}, {Tiwari},
  {Haney}, \& {Jetzer}}]{Grobner2020}
{Gr{\"o}bner}, M., {Ishibashi}, W., {Tiwari}, S., {Haney}, M., \& {Jetzer}, P.
  2020, \href{http://dx.doi.org/10.1051/0004-6361/202037681}{\aap},
  \href{https://ui.adsabs.harvard.edu/abs/2020A&A...638A.119G}{638, A119}

\bibitem[{{Hamers} {et~al.}(2018){Hamers}, {Bar-Or}, {Petrovich}, \&
  {Antonini}}]{Hamers2018}
{Hamers}, A.~S., {Bar-Or}, B., {Petrovich}, C., \& {Antonini}, F. 2018,
  \href{http://dx.doi.org/10.3847/1538-4357/aadae2}{\apj},
  \href{https://ui.adsabs.harvard.edu/abs/2018ApJ...865....2H}{865, 2}

\bibitem[{{Hunter}(2007)}]{Hunter2007}
{Hunter}, J.~D. 2007, \href{http://dx.doi.org/10.1109/MCSE.2007.55}{Computing
  in Science and Engineering},
  \href{https://ui.adsabs.harvard.edu/abs/2007CSE.....9...90H}{9, 90}

\bibitem[{{Ishibashi} \& {Gr{\"o}bner}(2020)}]{Ishibashi2020}
{Ishibashi}, W., \& {Gr{\"o}bner}, M. 2020,
  \href{http://dx.doi.org/10.1051/0004-6361/202037799}{\aap},
  \href{https://ui.adsabs.harvard.edu/abs/2020A&A...639A.108I}{639, A108}

\bibitem[{{Kremer} {et~al.}(2019){Kremer}, {Chatterjee}, {Ye}, {Rodriguez}, \&
  {Rasio}}]{Kremer2019}
{Kremer}, K., {Chatterjee}, S., {Ye}, C.~S., {Rodriguez}, C.~L., \& {Rasio},
  F.~A. 2019, \href{http://dx.doi.org/10.3847/1538-4357/aaf646}{\apj},
  \href{https://ui.adsabs.harvard.edu/abs/2019ApJ...871...38K}{871, 38}

\bibitem[{{Leigh} {et~al.}(2018){Leigh}, {Geller}, {McKernan}, {Ford}, {Mac
  Low}, {Bellovary}, {Haiman}, {Lyra}, {Samsing}, {O'Dowd}, {Kocsis}, \&
  {Endlich}}]{Leigh2018}
{Leigh}, N.~W.~C., {Geller}, A.~M., {McKernan}, B., {et~al.} 2018,
  \href{http://dx.doi.org/10.1093/mnras/stx3134}{\mnras},
  \href{https://ui.adsabs.harvard.edu/abs/2018MNRAS.474.5672L}{474, 5672}

\bibitem[{{Li} \& {Lai}(2020)}]{Li2020}
{Li}, J., \& {Lai}, D. 2020,
  \href{http://dx.doi.org/10.3847/2041-8213/aba2c4}{\apjl},
  \href{https://ui.adsabs.harvard.edu/abs/2020ApJ...898L..20L}{898, L20}

\bibitem[{{Li} \& {Lai}(2022)}]{Li2022}
{Li}, R., \& {Lai}, D. 2022,
  \href{https://ui.adsabs.harvard.edu/abs/2022arXiv220207633L}{arXiv e-prints,
  arXiv:2202.07633}

\bibitem[{{Li} {et~al.}(2021{\natexlab{a}}){Li}, {Dempsey}, {Li}, {Li}, \&
  {Li}}]{Li2021a}
{Li}, Y.-P., {Dempsey}, A.~M., {Li}, H., {Li}, S., \& {Li}, J.
  2021{\natexlab{a}},
  \href{https://ui.adsabs.harvard.edu/abs/2021arXiv211211057L}{arXiv e-prints,
  arXiv:2112.11057}

\bibitem[{{Li} {et~al.}(2021{\natexlab{b}}){Li}, {Dempsey}, {Li}, {Li}, \&
  {Li}}]{Li2021}
{Li}, Y.-P., {Dempsey}, A.~M., {Li}, S., {Li}, H., \& {Li}, J.
  2021{\natexlab{b}}, \href{http://dx.doi.org/10.3847/1538-4357/abed48}{\apj},
  \href{https://ui.adsabs.harvard.edu/abs/2021ApJ...911..124L}{911, 124}

\bibitem[{{Lipunov} {et~al.}(1997){Lipunov}, {Postnov}, \&
  {Prokhorov}}]{Lipunov1997}
{Lipunov}, V.~M., {Postnov}, K.~A., \& {Prokhorov}, M.~E. 1997, Astronomy
  Letters, \href{https://ui.adsabs.harvard.edu/abs/1997AstL...23..492L}{23,
  492}

\bibitem[{{Liu} \& {Lai}(2018)}]{Liu2018}
{Liu}, B., \& {Lai}, D. 2018,
  \href{http://dx.doi.org/10.3847/1538-4357/aad09f}{\apj},
  \href{https://ui.adsabs.harvard.edu/abs/2018ApJ...863...68L}{863, 68}

\bibitem[{{Liu} \& {Lai}(2019)}]{Liu2019}
---. 2019, \href{http://dx.doi.org/10.1093/mnras/sty3432}{\mnras},
  \href{https://ui.adsabs.harvard.edu/abs/2019MNRAS.483.4060L}{483, 4060}

\bibitem[{{Liu} \& {Lai}(2021)}]{Liu2021}
---. 2021, \href{http://dx.doi.org/10.1093/mnras/stab178}{\mnras},
  \href{https://ui.adsabs.harvard.edu/abs/2021MNRAS.502.2049L}{502, 2049}

\bibitem[{{Liu} {et~al.}(2019{\natexlab{a}}){Liu}, {Lai}, \& {Wang}}]{Liu2019a}
{Liu}, B., {Lai}, D., \& {Wang}, Y.-H. 2019{\natexlab{a}},
  \href{http://dx.doi.org/10.3847/1538-4357/ab2dfb}{\apj},
  \href{https://ui.adsabs.harvard.edu/abs/2019ApJ...881...41L}{881, 41}

\bibitem[{{Liu} {et~al.}(2019{\natexlab{b}}){Liu}, {Lai}, \& {Wang}}]{Liu2019b}
---. 2019{\natexlab{b}},
  \href{http://dx.doi.org/10.3847/2041-8213/ab40c0}{\apjl},
  \href{https://ui.adsabs.harvard.edu/abs/2019ApJ...883L...7L}{883, L7}

\bibitem[{{Mandel} \& {de Mink}(2016)}]{Mandel2016}
{Mandel}, I., \& {de Mink}, S.~E. 2016,
  \href{http://dx.doi.org/10.1093/mnras/stw379}{\mnras},
  \href{https://ui.adsabs.harvard.edu/abs/2016MNRAS.458.2634M}{458, 2634}

\bibitem[{{Marchant} {et~al.}(2016){Marchant}, {Langer}, {Podsiadlowski},
  {Tauris}, \& {Moriya}}]{Marchant2016}
{Marchant}, P., {Langer}, N., {Podsiadlowski}, P., {Tauris}, T.~M., \&
  {Moriya}, T.~J. 2016,
  \href{http://dx.doi.org/10.1051/0004-6361/201628133}{\aap},
  \href{https://ui.adsabs.harvard.edu/abs/2016A&A...588A..50M}{588, A50}

\bibitem[{{McKernan} {et~al.}(2014){McKernan}, {Ford}, {Kocsis}, {Lyra}, \&
  {Winter}}]{McKernan2014}
{McKernan}, B., {Ford}, K.~E.~S., {Kocsis}, B., {Lyra}, W., \& {Winter}, L.~M.
  2014, \href{http://dx.doi.org/10.1093/mnras/stu553}{\mnras},
  \href{https://ui.adsabs.harvard.edu/abs/2014MNRAS.441..900M}{441, 900}

\bibitem[{{McKernan} {et~al.}(2012){McKernan}, {Ford}, {Lyra}, \&
  {Perets}}]{McKernan2012}
{McKernan}, B., {Ford}, K.~E.~S., {Lyra}, W., \& {Perets}, H.~B. 2012,
  \href{http://dx.doi.org/10.1111/j.1365-2966.2012.21486.x}{\mnras},
  \href{https://ui.adsabs.harvard.edu/abs/2012MNRAS.425..460M}{425, 460}

\bibitem[{{McKernan} {et~al.}(2018){McKernan}, {Ford}, {Bellovary}, {Leigh},
  {Haiman}, {Kocsis}, {Lyra}, {Mac Low}, {Metzger}, {O'Dowd}, {Endlich}, \&
  {Rosen}}]{McKernan2018}
{McKernan}, B., {Ford}, K.~E.~S., {Bellovary}, J., {et~al.} 2018,
  \href{http://dx.doi.org/10.3847/1538-4357/aadae5}{\apj},
  \href{https://ui.adsabs.harvard.edu/abs/2018ApJ...866...66M}{866, 66}

\bibitem[{{McKernan} {et~al.}(2019){McKernan}, {Ford}, {Bartos}, {Graham},
  {Lyra}, {Marka}, {Marka}, {Ross}, {Stern}, \& {Yang}}]{McKernan2019}
{McKernan}, B., {Ford}, K.~E.~S., {Bartos}, I., {et~al.} 2019,
  \href{http://dx.doi.org/10.3847/2041-8213/ab4886}{\apjl},
  \href{https://ui.adsabs.harvard.edu/abs/2019ApJ...884L..50M}{884, L50}

\bibitem[{{Miller} \& {Hamilton}(2002)}]{Miller2002}
{Miller}, M.~C., \& {Hamilton}, D.~P. 2002,
  \href{http://dx.doi.org/10.1086/341788}{\apj},
  \href{https://ui.adsabs.harvard.edu/abs/2002ApJ...576..894M}{576, 894}

\bibitem[{{Miller} \& {Lauburg}(2009)}]{Miller2009}
{Miller}, M.~C., \& {Lauburg}, V.~M. 2009,
  \href{http://dx.doi.org/10.1088/0004-637X/692/1/917}{\apj},
  \href{https://ui.adsabs.harvard.edu/abs/2009ApJ...692..917M}{692, 917}

\bibitem[{{O'Leary} {et~al.}(2006){O'Leary}, {Rasio}, {Fregeau}, {Ivanova}, \&
  {O'Shaughnessy}}]{OLeary2006}
{O'Leary}, R.~M., {Rasio}, F.~A., {Fregeau}, J.~M., {Ivanova}, N., \&
  {O'Shaughnessy}, R. 2006, \href{http://dx.doi.org/10.1086/498446}{\apj},
  \href{https://ui.adsabs.harvard.edu/abs/2006ApJ...637..937O}{637, 937}

\bibitem[{{Peters}(1964)}]{Peters1964}
{Peters}, P.~C. 1964,
  \href{http://dx.doi.org/10.1103/PhysRev.136.B1224}{Physical Review},
  \href{https://ui.adsabs.harvard.edu/abs/1964PhRv..136.1224P}{136, 1224}

\bibitem[{{Petrovich} \& {Antonini}(2017)}]{Petrovich2017}
{Petrovich}, C., \& {Antonini}, F. 2017,
  \href{http://dx.doi.org/10.3847/1538-4357/aa8628}{\apj},
  \href{https://ui.adsabs.harvard.edu/abs/2017ApJ...846..146P}{846, 146}

\bibitem[{{Podsiadlowski} {et~al.}(2003){Podsiadlowski}, {Rappaport}, \&
  {Han}}]{Podsiadlowski2003}
{Podsiadlowski}, P., {Rappaport}, S., \& {Han}, Z. 2003,
  \href{http://dx.doi.org/10.1046/j.1365-8711.2003.06464.x}{\mnras},
  \href{https://ui.adsabs.harvard.edu/abs/2003MNRAS.341..385P}{341, 385}

\bibitem[{{Portegies Zwart} \& {McMillan}(2000)}]{PortegiesZwart2000}
{Portegies Zwart}, S.~F., \& {McMillan}, S. L.~W. 2000,
  \href{http://dx.doi.org/10.1086/312422}{\apjl},
  \href{https://ui.adsabs.harvard.edu/abs/2000ApJ...528L..17P}{528, L17}

\bibitem[{{Pu} \& {Lai}(2021)}]{Pu2021}
{Pu}, B., \& {Lai}, D. 2021,
  \href{http://dx.doi.org/10.1093/mnras/stab2504}{\mnras},
  \href{https://ui.adsabs.harvard.edu/abs/2021MNRAS.508..597P}{508, 597}

\bibitem[{{Quinlan} \& {Shapiro}(1989)}]{Quinlan1989}
{Quinlan}, G.~D., \& {Shapiro}, S.~L. 1989,
  \href{http://dx.doi.org/10.1086/167745}{\apj},
  \href{https://ui.adsabs.harvard.edu/abs/1989ApJ...343..725Q}{343, 725}

\bibitem[{{Rein} \& {Liu}(2012)}]{Rein2012}
{Rein}, H., \& {Liu}, S.~F. 2012,
  \href{http://dx.doi.org/10.1051/0004-6361/201118085}{\aap},
  \href{https://ui.adsabs.harvard.edu/abs/2012A&A...537A.128R}{537, A128}

\bibitem[{{Rein} \& {Spiegel}(2015)}]{Rein2015}
{Rein}, H., \& {Spiegel}, D.~S. 2015,
  \href{http://dx.doi.org/10.1093/mnras/stu2164}{\mnras},
  \href{https://ui.adsabs.harvard.edu/abs/2015MNRAS.446.1424R}{446, 1424}

\bibitem[{{Rodriguez} {et~al.}(2015){Rodriguez}, {Morscher}, {Pattabiraman},
  {Chatterjee}, {Haster}, \& {Rasio}}]{Rodriguez2015}
{Rodriguez}, C.~L., {Morscher}, M., {Pattabiraman}, B., {et~al.} 2015,
  \href{http://dx.doi.org/10.1103/PhysRevLett.115.051101}{\prl},
  \href{https://ui.adsabs.harvard.edu/abs/2015PhRvL.115e1101R}{115, 051101}

\bibitem[{{Samsing} \& {D'Orazio}(2018)}]{Samsing2018}
{Samsing}, J., \& {D'Orazio}, D.~J. 2018,
  \href{http://dx.doi.org/10.1093/mnras/sty2334}{\mnras},
  \href{https://ui.adsabs.harvard.edu/abs/2018MNRAS.481.5445S}{481, 5445}

\bibitem[{{Samsing} {et~al.}(2014){Samsing}, {MacLeod}, \&
  {Ramirez-Ruiz}}]{Samsing2014}
{Samsing}, J., {MacLeod}, M., \& {Ramirez-Ruiz}, E. 2014,
  \href{http://dx.doi.org/10.1088/0004-637X/784/1/71}{\apj},
  \href{https://ui.adsabs.harvard.edu/abs/2014ApJ...784...71S}{784, 71}

\bibitem[{{Samsing} {et~al.}(2020){Samsing}, {Bartos}, {D'Orazio}, {Haiman},
  {Kocsis}, {Leigh}, {Liu}, {Pessah}, \& {Tagawa}}]{Samsing2020}
{Samsing}, J., {Bartos}, I., {D'Orazio}, D.~J., {et~al.} 2020,
  \href{https://ui.adsabs.harvard.edu/abs/2020arXiv201009765S}{arXiv e-prints,
  arXiv:2010.09765}

\bibitem[{{Secunda} {et~al.}(2019){Secunda}, {Bellovary}, {Mac Low}, {Ford},
  {McKernan}, {Leigh}, {Lyra}, \& {S{\'a}ndor}}]{Secunda2019}
{Secunda}, A., {Bellovary}, J., {Mac Low}, M.-M., {et~al.} 2019,
  \href{http://dx.doi.org/10.3847/1538-4357/ab20ca}{\apj},
  \href{https://ui.adsabs.harvard.edu/abs/2019ApJ...878...85S}{878, 85}

\bibitem[{{Secunda} {et~al.}(2020){Secunda}, {Bellovary}, {Mac Low}, {Ford},
  {McKernan}, {Leigh}, {Lyra}, {S{\'a}ndor}, \& {Adorno}}]{Secunda2020}
---. 2020, \href{http://dx.doi.org/10.3847/1538-4357/abbc1d}{\apj},
  \href{https://ui.adsabs.harvard.edu/abs/2020ApJ...903..133S}{903, 133}

\bibitem[{{Silsbee} \& {Tremaine}(2017)}]{Silsbee2017}
{Silsbee}, K., \& {Tremaine}, S. 2017,
  \href{http://dx.doi.org/10.3847/1538-4357/aa5729}{\apj},
  \href{https://ui.adsabs.harvard.edu/abs/2017ApJ...836...39S}{836, 39}

\bibitem[{{Sirko} \& {Goodman}(2003)}]{Sirko2003}
{Sirko}, E., \& {Goodman}, J. 2003,
  \href{http://dx.doi.org/10.1046/j.1365-8711.2003.06431.x}{\mnras},
  \href{https://ui.adsabs.harvard.edu/abs/2003MNRAS.341..501S}{341, 501}

\bibitem[{{Smith} \& {Lissauer}(2009)}]{Smith2009}
{Smith}, A.~W., \& {Lissauer}, J.~J. 2009,
  \href{http://dx.doi.org/10.1016/j.icarus.2008.12.027}{\icarus},
  \href{https://ui.adsabs.harvard.edu/abs/2009Icar..201..381S}{201, 381}

\bibitem[{{Stone} {et~al.}(2017){Stone}, {Metzger}, \& {Haiman}}]{Stone2017}
{Stone}, N.~C., {Metzger}, B.~D., \& {Haiman}, Z. 2017,
  \href{http://dx.doi.org/10.1093/mnras/stw2260}{\mnras},
  \href{https://ui.adsabs.harvard.edu/abs/2017MNRAS.464..946S}{464, 946}

\bibitem[{{Tagawa} {et~al.}(2020{\natexlab{a}}){Tagawa}, {Haiman}, {Bartos}, \&
  {Kocsis}}]{Tagawa2020b}
{Tagawa}, H., {Haiman}, Z., {Bartos}, I., \& {Kocsis}, B. 2020{\natexlab{a}},
  \href{http://dx.doi.org/10.3847/1538-4357/aba2cc}{\apj},
  \href{https://ui.adsabs.harvard.edu/abs/2020ApJ...899...26T}{899, 26}

\bibitem[{{Tagawa} {et~al.}(2020{\natexlab{b}}){Tagawa}, {Haiman}, \&
  {Kocsis}}]{Tagawa2020a}
{Tagawa}, H., {Haiman}, Z., \& {Kocsis}, B. 2020{\natexlab{b}},
  \href{http://dx.doi.org/10.3847/1538-4357/ab9b8c}{\apj},
  \href{https://ui.adsabs.harvard.edu/abs/2020ApJ...898...25T}{898, 25}

\bibitem[{{Tanaka} \& {Ward}(2004)}]{Tanaka2004}
{Tanaka}, H., \& {Ward}, W.~R. 2004,
  \href{http://dx.doi.org/10.1086/380992}{\apj},
  \href{https://ui.adsabs.harvard.edu/abs/2004ApJ...602..388T}{602, 388}

\bibitem[{{The LIGO Scientific Collaboration} {et~al.}(2021){The LIGO
  Scientific Collaboration}, {the Virgo Collaboration}, {Abbott}, {Abbott},
  {Acernese}, {Ackley}, {Adams}, {Adhikari}, {Adhikari}, {Adya}, \&
  et~al.}]{LIGO2021b}
{The LIGO Scientific Collaboration}, {the Virgo Collaboration}, {Abbott}, R.,
  {et~al.} 2021,
  \href{https://ui.adsabs.harvard.edu/abs/2021arXiv210801045T}{arXiv e-prints,
  arXiv:2108.01045}

\bibitem[{{Thompson} {et~al.}(2005){Thompson}, {Quataert}, \&
  {Murray}}]{Thompson2005}
{Thompson}, T.~A., {Quataert}, E., \& {Murray}, N. 2005,
  \href{http://dx.doi.org/10.1086/431923}{\apj},
  \href{https://ui.adsabs.harvard.edu/abs/2005ApJ...630..167T}{630, 167}

\bibitem[{{van der Walt} {et~al.}(2011){van der Walt}, {Colbert}, \&
  {Varoquaux}}]{Walt2011}
{van der Walt}, S., {Colbert}, S.~C., \& {Varoquaux}, G. 2011,
  \href{http://dx.doi.org/10.1109/MCSE.2011.37}{Computing in Science and
  Engineering},
  \href{https://ui.adsabs.harvard.edu/abs/2011CSE....13b..22V}{13, 22}

\bibitem[{{VanLandingham} {et~al.}(2016){VanLandingham}, {Miller}, {Hamilton},
  \& {Richardson}}]{VanLandingham2016}
{VanLandingham}, J.~H., {Miller}, M.~C., {Hamilton}, D.~P., \& {Richardson},
  D.~C. 2016, \href{http://dx.doi.org/10.3847/0004-637X/828/2/77}{\apj},
  \href{https://ui.adsabs.harvard.edu/abs/2016ApJ...828...77V}{828, 77}

\bibitem[{{Virtanen} {et~al.}(2020){Virtanen}, {Gommers}, {Oliphant},
  {Haberland}, {Reddy}, {Cournapeau}, {Burovski}, {Peterson}, {Weckesser},
  {Bright}, {van der Walt}, {Brett}, {Wilson}, {Millman}, {Mayorov}, {Nelson},
  {Jones}, {Kern}, {Larson}, {Carey}, {Polat}, {Feng}, {Moore}, {VanderPlas},
  {Laxalde}, {Perktold}, {Cimrman}, {Henriksen}, {Quintero}, {Harris},
  {Archibald}, {Ribeiro}, {Pedregosa}, {van Mulbregt}, \& {SciPy 1. 0
  Contributors}}]{Virtanen2020}
{Virtanen}, P., {Gommers}, R., {Oliphant}, T.~E., {et~al.} 2020,
  \href{http://dx.doi.org/10.1038/s41592-019-0686-2}{Nature Methods},
  \href{https://ui.adsabs.harvard.edu/abs/2020NatMe..17..261V}{17, 261}

\bibitem[{{Yang} {et~al.}(2019{\natexlab{a}}){Yang}, {Bartos}, {Haiman},
  {Kocsis}, {M{\'a}rka}, {Stone}, \& {M{\'a}rka}}]{Yang2019a}
{Yang}, Y., {Bartos}, I., {Haiman}, Z., {et~al.} 2019{\natexlab{a}},
  \href{http://dx.doi.org/10.3847/1538-4357/ab16e3}{\apj},
  \href{https://ui.adsabs.harvard.edu/abs/2019ApJ...876..122Y}{876, 122}

\bibitem[{{Yang} {et~al.}(2019{\natexlab{b}}){Yang}, {Bartos}, {Gayathri},
  {Ford}, {Haiman}, {Klimenko}, {Kocsis}, {M{\'a}rka}, {M{\'a}rka}, {McKernan},
  \& {O'Shaughnessy}}]{Yang2019b}
{Yang}, Y., {Bartos}, I., {Gayathri}, V., {et~al.} 2019{\natexlab{b}},
  \href{http://dx.doi.org/10.1103/PhysRevLett.123.181101}{\prl},
  \href{https://ui.adsabs.harvard.edu/abs/2019PhRvL.123r1101Y}{123, 181101}

\bibitem[{{Ziosi} {et~al.}(2014){Ziosi}, {Mapelli}, {Branchesi}, \&
  {Tormen}}]{Ziosi2014}
{Ziosi}, B.~M., {Mapelli}, M., {Branchesi}, M., \& {Tormen}, G. 2014,
  \href{http://dx.doi.org/10.1093/mnras/stu824}{\mnras},
  \href{https://ui.adsabs.harvard.edu/abs/2014MNRAS.441.3703Z}{441, 3703}
\end{thebibliography}

\end{document}